\documentclass{aa}
\usepackage{amsmath}
\usepackage{graphicx}
\usepackage{changebar}
\usepackage{natbib}
\usepackage{subfigure}
\usepackage{multirow}
\usepackage{hyperref}

\newcommand{\kms}{\mbox{km s$^{-1}$}}

\newcommand{\vlsr}{V$_{\rm{LSR}}$}
\newcommand{\lsun}{L$_\odot$}
\newcommand{\msun}{M$_\odot$}
\begin{document}
\title{Radio continuum and molecular line observations of four bright-rimmed clouds}
\author{J. S. Urquhart\inst{1}, M. A. Thompson\inst{2}, L. K. Morgan\inst{3,4} \& Glenn J. White\inst{4,5}}
\offprints{J. S. Urquhart: jsu@ast.leeds.ac.uk}
\institute{
Department of Physics and Astronomy, University of Leeds, Leeds, LS2 9JT, UK
\and
 School of Physics Astronomy \& Maths, University of Hertfordshire, College Lane, Hatfield, AL10 9AB, UK
\and
Centre for Astrophysics and Planetary Science, School of Physical Sciences, University of 
Kent, Canterbury, CT2 7NR, UK
\and
Green Bank Telescope, P.O. Box 2, Green Bank, WV 24944, USA
\and 
Dept. of Physics \& Astronomy, The Open University, Walton Hall, Milton
Keynes, MK7 6AA, UK
\and
Space Physics Division, Space Science \& Technology Division, CCLRC
Rutherford Appleton Laboratory, Chilton, Didcot, Oxfordshire, OX11 0QX,
UK
}
\date{}
\abstract{}{To search for evidence of triggered star formation within four bright-rimmed clouds, SFO~58, SFO~68, SFO~75 and
SFO~76.}{We present the results of radio continuum and molecular line observations conducted using the Mopra millimetre-wave telescope and Australia Telescope Compact Array. We use the \mbox{$J$=1--0} transitions of $^{12}$CO, $^{13}$CO and C$^{18}$O to trace the distribution of molecular material and to study its kinematics.}{These observations reveal the presence of a dense core ($n_{\rm{H}_2}>10^4$~cm$^{-3}$) embedded within each cloud, and the presence of a layer of hot ionised gas coincided with their bright-rims. The ionised gas has electron densities significantly higher than the critical density ($>$~25 cm$^{-3}$) above which an ionised boundary layer can form and be maintained, strongly supporting the hypothesis that these clouds are being photoionised by the nearby OB star(s). Using a simple pressure-based argument, photoionisation is shown to have a profound effect on the stability of these cores, leaving SFO~58 and SFO~68 on the edge of gravitational stability, and is also likely to have rendered SFO~75 and SFO~76 unstable to gravitational collapse. From an evaluation of the pressure balance between the ionised and molecular gas, SFO~58 and SFO~68 are identified as being in a post-pressure balance state, while SFO~75 and SFO~76 are more likely to be in a pre-pressure balance state. We find secondary evidence for the presence of ongoing star formation within SFO~58 and SFO~68, such as molecular outflows, OH, H$_2$O and methanol masers, and identify a potential embedded UC HII region, but find no evidence for any ongoing star formation within SFO~75 and SFO~76.}{Our results are consistent with the star formation within SFO~58 and SFO~68 having been triggered by the radiatively driven implosion of these clouds.}
	
\authorrunning{J. S. Urquhart et al.}
\titlerunning{Radio observations of four bright-rimmed clouds}

\maketitle

\section{Introduction}

From the moment they turn on OB stars begin to drive an ionisation front into the surrounding molecular material, photo-evaporating and dissipating the molecular cloud from which they have formed. The rapidly expanding HII region leads to the formation of a dense shell of neutral gas, swept up in front of the ionisation front. These dense shells, and dense neutral clumps of material, surrounding evolved HII regions have long been suspected to be regions where star formation could have been triggered \mbox{(\citealt{elmegreen1977,sandford1982})}. Bright-Rimmed Clouds (BRCs) are small molecular clouds located on the edges of evolved HII regions and are considered, due to their relatively simple geometry and isolation within HII regions, to be ideal laboratories in which to study the physical processes involved in triggered star formation.

The photoionisation of the BRCs surface layers by UV photons from nearby OB stars leads to the formation of a layer of hot ionised gas, known as an \emph{Ionised Boundary Layer} (IBL), which surrounds the rim of the molecular cloud. The hot ionised gas streams off the surface of the cloud into the low density HII regions, resulting in a continuous mass loss by the cloud, known as a \emph{photo-evaporative flow} (\citealt{megeath1997}). Within the IBL the incoming ionising photon flux is balanced by recombination, with only a small fraction of ionising photons penetrating the IBL to ionise new material (\citealt{lefloch1994}), which replenishes the ionised material within the IBL lost to the photo-evaporative flow. If the IBL is over-pressured with respect to the molecular gas within the BRC, shocks are driven into the molecular gas, resulting in the compression of the cloud, and can lead to the formation of dense cores which are then triggered to collapse by the same (or a subsequent) shock front (\citealt{elmegreen1992}). The propagating shock front may also serve to trigger the collapse of pre-existing dense cores, thus leading to the creation of a new generation of stars. This method of triggered star formation is known as \emph{Radiative--Driven Implosion} (RDI) and may be responsible for the production of hundreds of stars in each HII region (\citealt{ogura2002}), and perhaps even contributing up to $\sim 15$ \% of the low-to-intermediate mass IMF (\citealt{sugitani2000}). 

Shocks continue to be driven into the cloud, compressing the molecular material until the internal density and
pressure is balanced with the pressure of the IBL; after which the shock fronts dissipate and the cloud is
considered to be in a quasi-steady state known as the cometary stage (\citealt{bertoldi1990,lefloch1994}). Once
equilibrium is reached the ionisation front is unable to have any further influence on the internal dynamics of the
cloud, but continues to propagate into the cloud photo-evaporating the molecular gas, which streams into the HII
region, eroding the cloud and accelerating it radially away from the ionising star via the \emph{rocket effect}
(\citealt{oort1955}). The mass loss resulting from photoionisation ultimately leads to the destruction of the
cloud on a timescale of several million years.

A search of the IRAS point source catalogue, correlated with the Sharpless HII region catalogue (\citealt{sharpless1959}) and the ESO(R) Southern Hemisphere Atlas resulted in a total of 89 BRCs being identified with an associated IRAS point source, 44 in the northern and 45 in the southern sky (\citealt{sugitani1991, sugitani1994}; collectively known as the SFO catalogue). The association of an IRAS point source located within these BRCs suggests that these clouds might contain embedded protostars. Several comprehensive studies of individual BRCs have been reported (e.g. \citealt{lefloch1997, megeath1997,yamaguchi1999,codella2001, devries2002,dobashi2002,thompson2004a}), all of which have confirmed their association with protostellar cores. However, the question of star formation occurring more widely in these objects still remains unclear, and evidence for triggered star formation within the small number of BRCs so far observed remains circumstantial and inconclusive. To address these issues we are currently conducting a complete census of the SFO catalogue; here we report the results of a detailed study of four southern BRCs.

In an earlier paper (Thompson, Urquhart \& White, 2004b; hereafter Paper~I) we reported the results of a relatively
low angular resolution radio
continuum survey of the 45 southern BRCs taken from the SFO catalogue. In that survey we detected radio continuum
emission toward eighteen BRCs. In each case the radio emission was found to correlate extremely well with both the
morphology of the optical bright rim seen in the DSS~(R band) image and the Photo-Dominated Region (PDR) traced by
the Midcourse Space eXperiment (MSX)\footnote{Available from the NASA/IPAC Infrared Processing and Analysis Center
and NASA/IPAC Infrared Space Archive both held at http://www.ipac.caltech.edu.} 8~$\mu$m image
(\citealt{price2001}), consistent with the hypothesis  that these eighteen BRCs are being photoionised by the
nearby OB star(s).

In this paper we present high resolution molecular line and radio continuum observations toward four clouds (i.e. SFO~58, SFO~68, SFO~75 and SFO~76) from the eighteen photoionised clouds identified in Paper~I which displayed the best correlation between the optical, MIR and radio emission and appeared to possess the simplest geometry.  These clouds are thus considered ideal candidates for further investigation into the applicability of the RDI star-formation mechanism. From these observations we investigate the internal and external structure of these clouds and calculate their physical parameters, which when combined with archival data will provide a comprehensive picture of star formation within these BRCs and allow us to determine whether or not it could have been triggered. 

The structure of this paper is as follows: in Section~2 we briefly describe the morphologies of the BRCs and the HII regions in which they are located, including a summary of any previous observations that have been reported toward them. In Section~3 we describe our observation strategy and the molecular line and radio continuum observations, followed in Section~4 by the observational results and analyses. In Section~5 we  discuss the impact the ionisation front has had on the stability, morphology and future evolution of these clouds and try to evaluate the current state of star formation within each cloud and whether it could have been triggered. We  present a summary and our conclusions in Section~6. 

\section{Description of individual BRCs}
%\section{Summary of physical parameters}
\label{sect:summary_clouds}

\begin{figure*}
\begin{center}
\includegraphics[width=0.45\textwidth,trim=10 0 10 10]{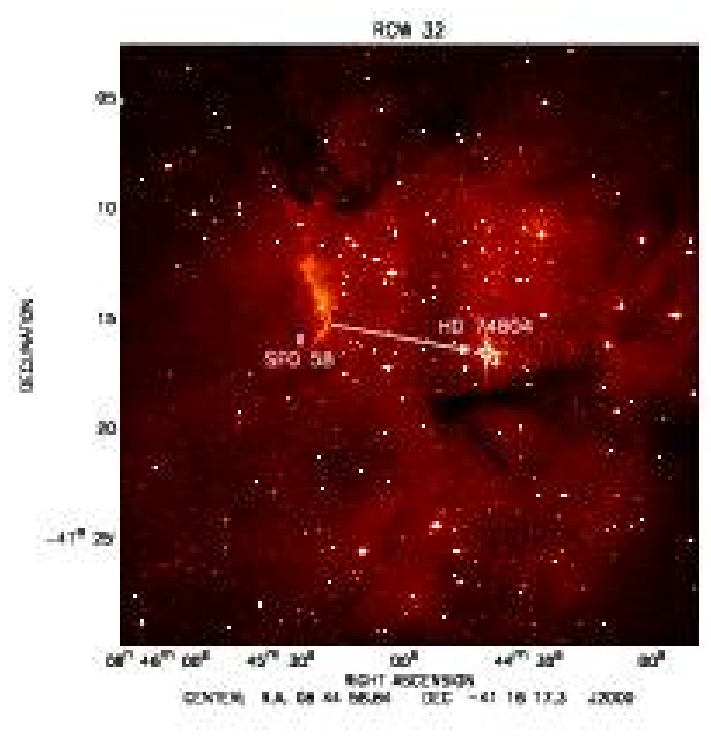}
\includegraphics[width=0.45\textwidth,trim=10 0 10 10]{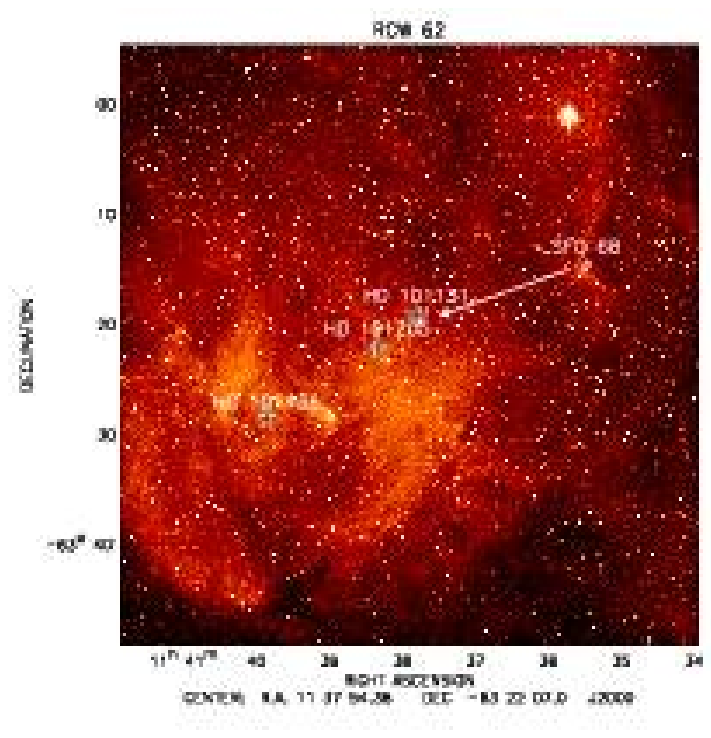}
\includegraphics[width=0.45\textwidth,trim=10 0 10 10]{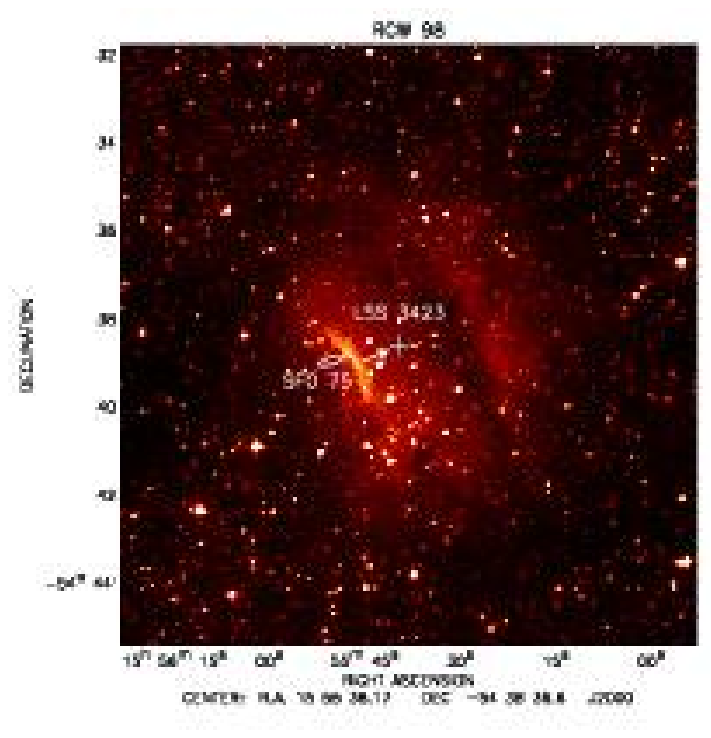}
\includegraphics[width=0.45\textwidth,trim=10 0 10 10]{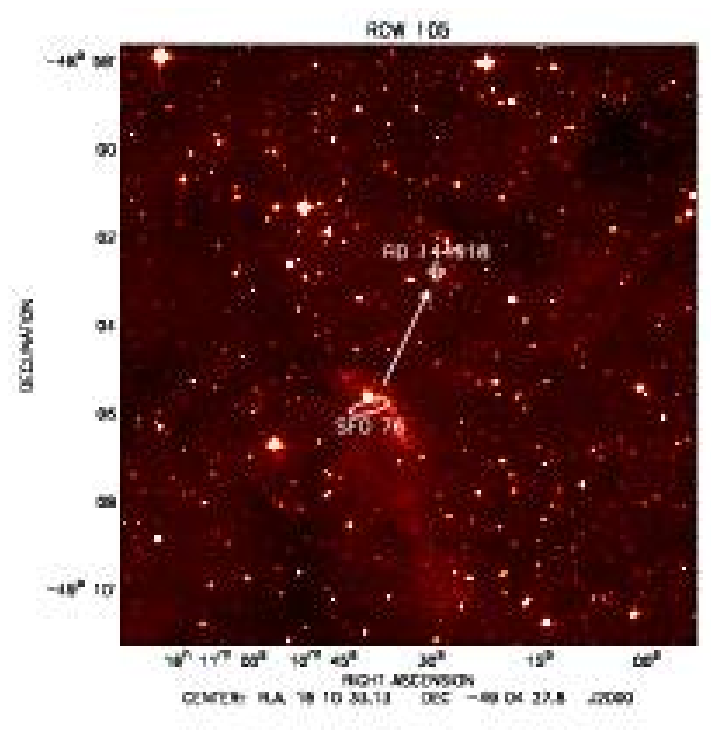}

\caption[DSS (R band) images of the four BRCs with IBLs]{DSS (R band) images of the four HII regions in which the SFO bright-rimmed clouds are located. The crosses indicate the position of the ionising star(s) as identified by \citet{yamaguchi1999}. The IRAS error ellipse is plotted in  white to indicate the position of the IRAS point source associated with each cloud (in the case of SFO~68 the size of the ellipse is twice the IRAS values). The arrow points from the rim of each cloud toward the ionising star(s).}
\label{fig:dss_brc_images}
\end{center}
\end{figure*}

In Figure~\ref{fig:dss_brc_images} we present a large scale DSS image of each BRC and the HII region in which they
are situated. In these we have indicated the position of the ionising star(s) as identified by
\citet{yamaguchi1999} and the IRAS point source with a
cross and an ellipse respectively. In these images it would appear that these BRCs are dense condensations of
material that are connected to an extended shell of molecular gas which surrounds the HII region that are beginning
to protrude into the HII region as the ionisation front ionises the less dense material around them. An arrow has
been added from the centre of each bright rim which points directly toward the ionising star(s). These images
clearly show the cloud morphologies to be curved in the general direction of the ionising stars; this is especially
evident for SFO~58 and SFO~68. The physical parameters relating to each HII region are presented in
Table~\ref{tbl:HII_regions}; the distances and  spectral types of the ionising star(s) have been taken from
\citet{yamaguchi1999}, the age of each HII region has been estimated by calculating their expansion timescale as
described by \citet{thompson2004a}.

\begin{table*}
\begin{center}
\caption[HII regions containing selected BRCs]{HII regions containing selected bright-rimmed clouds.}
\begin{tabular}{lccccc}
\hline
\hline
HII   & Distance  & Associated  & Ionising  & Spectral & HII region \\
region & (kpc)& BRC & star(s)& type & age (Myr)\\
\hline
RCW 32   & 0.7      & SFO~58    & HD 74804     & B0 V   & 0.26   \\
%BBW 347   & 2.7       & SFO 64     & LSS 2231    & B0 V     \\ 
RCW 62   & 1.7      &  SFO~68 & HD 101131    & O6.5 N  & 1.20  \\
     &         &      & HD 101205    & O6.5      \\
     &         &       & HD 101436    & O7.5      \\
RCW 98   & 2.8      & SFO~75    & LSS3423    & O9.5 IV   &0.31  \\
RCW 105   & 1.8      & SFO~76    & HD 144918    & O7     & 0.73 \\
\hline
\end{tabular}
\label{tbl:HII_regions}
\end{center}
\end{table*}

\begin{table*}[!ht]
\begin{center}
\caption[Parameters of the embedded IRAS point sources]{Parameters of the embedded IRAS point sources. (Upper limits are indicated by parenthesis.)}
\begin{tabular}{cccccc}
\hline
\hline
Cloud id. &IRAS id. &IRAS colour type  & Luminosity (\lsun) & Spectral type & Mass (\msun) \\
\hline 
SFO~58   & 08435--4105    & hot cirrus     & 140     &  B7 V  & 4.2   \\
%SFO 64  & 11101--5829      & class 0/UC HII & 14000    &  B0.5  & 15.9  \\ 
%SFO 67   & 11317--6254      & hot cirrus  & (470)    & B5    &6.0\\
SFO~68   & 11332--6258   & class 0/UC HII      & (3400)   &  (B1--B2)  &(10.6)  \\
SFO~75  & 15519--5430   & hot cirrus      & 34000    &  O9.5 & 20.6    \\
SFO~76   & 16069--4858   & class 0/UC HII     & 5600   &  B1 & 12.2  \\

\hline

\end{tabular}
\label{tbl:IRAS_sources}
\end{center}
\end{table*}

The IRAS point sources can be clearly seen to lie within the cloud, slightly behind the rim with respect to the
direction of ionisation. The parameters of the IRAS point sources associated with each BRC are presented in
Table~\ref{tbl:IRAS_sources}; the luminosities and classifications of the IRAS colours have been taken from
\citet{sugitani1994}, the spectral types of the embedded IRAS sources have been estimated using the tables of
\citet{panagia1973} and \citet{de_jager1987} assuming that the IRAS infrared luminosity is due to the presence of a
single Zero Age Main Sequence (ZAMS) star, the masses have been estimated using the \mbox{L$_{\rm{star}}$ $\simeq$
M$_{\rm{star}}^{3.45}$} relationship (\citealt{allen1973}).  Although the assumption that the infrared luminosity
of each IRAS point source is due to a single embedded star is a rather crude approximation, it does enable an upper
limit to the spectral type of the embedded star to be estimated.

There are a couple of interesting points to note from Table~\ref{tbl:IRAS_sources}. Firstly, three of the four BRCs could possibly harbour Massive Young Stellar Objects (MYSOs), and secondly, two BRCs have IRAS colours consistent with the presence of UC HII regions.

\subsection{SFO~58}

The BRC SFO~58 is situated on the edge of the HII region RCW~32, which is located at a heliocentric distance of $\sim$~700 pc (\citealt{georgelin1973}). The ionising star of RCW~32 has been identified as HD 74804, a B0 V--B4 II star (\citealt{yamaguchi1999}, and references therein), which is located at a projected distance of 1.53~pc from the bright rim of SFO~58.

\subsection{SFO~68}

SFO~68 is located on the northwestern edge of RCW~62, a bright HII region located approximately \mbox{1.7 kpc} from the Sun (\citealt{yamaguchi1999}). 
The HII region is driven by three O stars, HD~101131, HD~101205 and HD~101436, which have the spectral types of O6.5, O6.5 and O7.5 respectively (\citealt{yamaguchi1999}). At \mbox{1.7 kpc} the projected distances between SFO~68 and the three ionising stars range between 7.4--14.9 pc, with HD 101131 providing the vast majority of the ionising flux ($>90$ \%) impinging upon the surface of SFO~68.

SFO~68 has IRAS colours consistent with that of an UC~HII region, which has led to SFO~68 being included in several surveys of high-mass star forming regions, such as maser surveys (\citealt{braz1989,macleod1992,caswell1995}), and molecular line surveys (\citealt{zinchenko1995,bronfman1996}). H$_2$O, OH (\citealt{braz1989}) and 6.7~GHz methanol masers (\citealt{macleod1992,caswell1995}) have all been detected toward SFO~68, suggesting the presence of ongoing high-mass star formation within SFO~68. This is supported by the FIR luminosity, from which the presence of a B1--B2~ZAMS star embedded within the cloud can be inferred. The \vlsr~of the detected maser emission range between $-$12 and $-$17~\kms, which is in good agreement with the \vlsr~obtained from our CO observations ($-$16.7~\kms; see Section~4.1). 

In addition to the detected masers, a survey of candidate UC HII regions conducted by \citet{bronfman1996} detected CS(\emph{J}=2--1) emission toward the IRAS point source embedded within SFO~68. CS is a high density tracer with a critical excitation density threshold between 10$^4$--10$^5$~cm$^{-3}$, and therefore confirms the presence of dense gas coincident with the position of the IRAS point source. This CS emission has a \vlsr~of $-$15.4 \kms; similar to our CO velocity, and the velocities measured from the masers, confirms they are dynamically associated with this cloud. \citet{zinchenko1995} mapped SFO~68 using the  \mbox{CS(\emph{J}=2--1)} transition as well as making single pointing observations toward the molecular peak, identified from the CS map, in the C$^{34}$S(\emph{J}=2--1) and $^{12}$CO(\emph{J}=1--0) molecular transition lines. \citet{zinchenko1995} estimated the main beam temperature and mass of the molecular cloud to be $\sim$~24~K and 425~\msun~respectively. 

\subsection{SFO~75}
The bright-rimmed cloud SFO~75 is located on the southeastern edge of the HII region RCW~98. Embedded within SFO~75 is the IRAS point source 15519--5430, which is the most luminous in the SFO catalogue with an FIR luminosity, \emph{L}$_{\rm{FIR}}$ $\sim$ $3.4\times10^{4}$~L$_\odot$. The surface exposed to the HII region is being ionised by LSS 3423, an O9.5~IV star located 0.61~pc to the northwest. RCW~98 lies at a heliocentric distance of 2.8~kpc (\citealt{yamaguchi1999}).

\subsection{SFO~76}

The exciting star of RCW~105 is HD~144918, an O6 star located at a projected distance of 1.79~pc to the northwest of SFO~76 assuming a heliocentric distance of 1.8~kpc (\citealt{yamaguchi1999}). The IRAS point source 16069--4858 is located very close to the rim of the BRC and has an FIR luminosity of 5600~\lsun, and colours consistent with the presence of an UC HII region, which has led to this source being included in many of the same surveys of high-mass star forming regions as SFO~68. However, unlike SFO~68, searches for H$_2$O, OH (\citealt{braz1989,caswell1995}) and 6.7~GHz methanol masers (\citealt{walsh1997}) resulted in non-detections. CS emission has been detected toward SFO~76 (\citealt{bronfman1996}), confirming the presence of dense molecular gas coincident with the position of the IRAS point source. The FIR luminosity is consistent with the presence of a single B1 ZAMS star, supporting the identification of this cloud as a high-mass star forming region. However, the non-detection of maser emission suggests that either SFO~76 is at an earlier stage of development than SFO~68, or that it is simply in the process of forming a cluster of intermediate-mass stars.

\section{Observations and data reduction}
\subsection{Survey strategy}

Theoretical RDI models \citep{bertoldi1989,bertoldi1990,lefloch1994} suggest that the pressure balance between the IBL and the molecular gas can be used to identify clouds in which star formation may have been triggered, or is likely to be triggered in the future. A comparison of the internal and external pressures can result in three possible scenarios depending on whether the cloud is  (1) over-pressured, (2) under-pressured or (3) in approximate pressure balance with the respect to the IBL. The  implications of each of these scenarios are as follows:
 
\begin{enumerate}

\item Over-pressured clouds: the ionisation front is likely to have stalled at the surface of these clouds where it
will remain, unable to overcome the internal pressure of the cloud until the pressure in the IBL increases to match
that of the molecular cloud. The ionisation front has no dynamical effect on
these clouds and therefore is unlikely to have influenced any current,  or future, star formation within these
clouds.

\item Under-pressured clouds: these clouds are thought to have only recently been exposed to the ionisation front, and although it is highly likely that shocks are currently being driven into these clouds, these shocks have not yet led to the equalisation of the internal and external pressures, leaving the ionisation front stalled at the surface of these clouds. These clouds are thought to be in a \emph{pre-pressure balance state}. Any current star formation present in these clouds is unlikely to have been triggered and is more likely to be pre-existing.

\item Approximate pressure balance: these clouds are thought to have been initially under-pressured with respect to the ionisation front, however, shocks have propagated through the surface layers compressing the molecular gas, leading to an equalisation of the internal and external pressures, and leaving a dense core in its wake as it continues toward the rear of the cloud. These clouds are said to be in a \emph{post-pressure balance state}. Models suggest that the star formation within these clouds may have been triggered (\citealt{lefloch1994}).

\end{enumerate}

In order to evaluate the state of the pressure balance the physical properties of the ionised and molecular gas need to be measured. A combination of radio continuum and molecular line observations have previously been successfully used to determine the current state of several clouds (e.g. \citealt{lefloch1997,white1999,lefloch2002,thompson2004a}). To build on the models, and these previous observational studies, we have made high resolution molecular line and radio continuum observations of four BRCs.

\subsection{CO observations}
\label{sect:co_observations}

Observations of the four BRCs were made during June 2003 in the $J$=1--0 rotational lines of $^{12}$CO, $^{13}$CO and C$^{18}$O using the Mopra millimetre-wave telescope. Mopra is a 22 metre telescope located near Coonabarabran, New South Wales, Australia.\footnote{Mopra is operated by the Australia Telescope National Facility, CSIRO and the University of New South Wales.} The telescope is situated at an elevation of 866 metres above sea level, and  at a latitude of 31 degrees south. 

The receiver is a cryogenically cooled \mbox{($\sim$ 4 K)}, low-noise, Superconductor-Insulator-Superconductor (SIS) junction mixer with a frequency range between 85--116 GHz, corresponding to a half-power beam-width of \mbox{36--33 $\pm$ 2\arcsec} (Mopra Technical Summary version 10).\footnote{Available at http://www.narrabri.atnf.csiro.au/mopra/.} The receiver can be tuned to either single or double side-band mode. The incoming signal is separated into two channels, using a polarisation splitter, each of which can be tuned separately allowing two channels to be observed simultaneously. The receiver backend is a digital autocorrelator capable of providing two simultaneous outputs with an instantaneous bandwidth between 4--256 MHz.

For these observations a bandwidth of 64 MHz with a 1024-channel digital autocorrelator was used, giving a frequency resolution of \mbox{62.5 kHz} and a
velocity resolution of \mbox{0.16--0.17 km s$^{-1}$} over the \mbox{109--115 GHz} frequency range. For the $^{12}$CO and $^{13}$CO observations the second channel was tuned to \mbox{86.2 GHz} (SiO maser frequency) to allow pointing corrections to be performed during the observations. However, both bands were tuned to \mbox{109.782 GHz} for the C$^{18}$O observations in order to optimise the signal-to-noise ratio. System temperatures were between  $\sim 500$--$600$ K for $^{12}$CO and $\sim 250$--$350$~K for both $^{13}$CO and C$^{18}$O depending on weather conditions and telescope elevation, but were found to be stable over the short time periods required to complete each map, varying by no more than approximately 10 \%.\footnote{With the exception of the $^{12}$CO emission observed toward SFO~58 which suffered from large system temperature variation due to poor weather rendering the $^{12}$CO map unreliable. (Only the $^{13}$CO map will be presented for this source.)} Position-switching was used to subtract sky emission. Antenna pointing checks every two hours showed that the average pointing accuracy was better than 10$^{\prime\prime}$ r.m.s..

\begin{table}[!ht]
\begin{center}
\caption[Summary of Mopra CO observations]{Summary of Mopra CO observations.}
\label{tbl:co_observations}
\begin{tabular}{lcccc}
\hline
\hline
Isotope  & Frequency & Velocity res. & Grid & Integration  \\
(\emph{J}=1--0)  & (GHz)   & (km s$^{-1}$) & size& time (s)\\
\hline
$^{12}$CO & 115.271 & 0.162   & $9\times9$ & 30 \\
$^{13}$CO& 110.201 & 0.170  &$9\times9$ & 30\\
C$^{18}$O& 109.782 & 0.170  &$3\times3$  & 120  \\
\hline
\end{tabular}
\label{tbl:co_line}
\end{center}
\end{table}

The $^{12}$CO and $^{13}$CO observations consisted of spectra taken of a $9\times9$ pixel grid centred on the cometary head of each cloud, using a grid spacing of 15\arcsec. Each grid position was observed for 30 seconds, interleaved with observations at an off-source reference position for 90 seconds after each row of 9 points. The C$^{18}$O maps consist of a smaller grid of $3\times3$ points with the same spacing, and were centred on the molecular peaks identified from the $^{13}$CO maps.  For each of the C$^{18}$O grid positions a total integration time of 2 minutes was used. A summary of the observational parameters is presented in Table~\ref{tbl:co_line} with the grid centres and off-source reference positions presented in Table~\ref{tbl:positions}. 

\begin{table*}
\begin{center}
\caption[Pointing centres and off-source reference positions]{Pointing centres and off-source reference positions for all four BRCs.}
\label{tbl:positions}
\begin{tabular}{cccccc}
\hline
\hline
Cloud  & IRAS &\multicolumn{2}{c}{Pointing centre} & \multicolumn{2}{c}{Reference position} \\
id.& id.& $\alpha$(2000) & $\delta$(2000) & $\alpha$(2000) & $\delta$(2000)  \\
\hline
SFO~58 & 08435-4105& 08:45:25.4 & $-$41:16:02 & 08:45:26.1 & $-$41:15:10 \\
%SFO 64 & 11101-5829 & 11:12:18.1 & $-$58:46:20 & 11:15:58.2& $-$58:27:29& BBW 347\\
%SFO 67 & 11317-6254& 11:34:00.7 & $-$63:11:19 & 11:28:44.7 & $-$63:49:56 & RCW 62 \\
SFO~68 & 11332-6258 & 11:35:31.9 & $-$63:14:51 & 11:24:20.5 & $-$64:09:56  \\
SFO~75 & 15519-5430 & 15:55:50.4 & $-$54:38:58 & 16:02:17.6 & $-$55:19:12 \\
SFO~76 & 16069-4858 & 16:10:38.6 & $-$49:05:52 & 16:04:09.3 & $-$48:48:24 \\
\hline
\end{tabular}

\end{center}
\end{table*}

The measured antenna temperatures, $T_A^*$, were corrected for atmospheric absorption, ohmic losses and rearward
spillover, by taking measurements of an ambient load (assumed to be at 290 K) placed in front of the receiver
following the method of \citet{kutner1981}. To correct for forward spillover and scattering, these data are
converted to the corrected receiver temperature scale, T$_R^*$, by taking account of the main beam efficiency,
\emph{B$_{eff}/F_{eff}$} $\sim0.42\pm0.02.$\footnote{Main beam efficiencies have only been accurately determined at
86, 100 and 115 GHz which have efficiencies of 0.49$\pm0.03$, 0.44$\pm0.03$ and 0.42$\pm0.02$ respectively
(\citealt{ladd2005}). Interpolating from these efficiencies it is easy to show that, although the efficiency is
probably not the same at 110 GHz (i.e. $^{13}$CO and  C$^{18}$O ) to that at 115 GHz, any difference is smaller
that the errors involved. We have therefore adopted the 115 GHz efficiency for all three CO lines.} All of the BRCs
have angular diameters larger than $\sim$ 80$^{\prime\prime}$, and therefore to take account of the contribution
made to the measured intensity from extended material that couples to the error beam, we have used the
extended beam efficiency (i.e. $\eta_{xb}\sim0.55$) to correct the $^{12}$CO measurements. Absolute calibration was
performed by comparing measured line temperatures of Orion~KL and M17SW to standard values. We estimate the
combined calibration uncertainties to be no more than 10 \%.

The ATNF data reduction package, SPC, was used to process the individual spectra. Sky-subtracted spectra were obtained by subtracting emission from the off-source reference position from the on-source data. A correction was made to account for the change in the shape of the dish as a function of elevation. The data have been Hanning-smoothed to improve the signal-to-noise ratio, reducing the velocity resolution to 0.32--0.34 \kms. 

%The line profiles toward the four BRCs are all relatively simple consisting of a single emission peak between the $-$80 to 80 \kms~velocity range.

\subsection{Radio observations}

Centimetre-wave continuum observations were carried out with the Australia Telescope Compact Array (ATCA)\footnote{The Australia Telescope Compact Array is funded by the Commonwealth of Australia for operation as a National Facility managed by CSIRO.} between June 2002 and September 2003. ATCA is located at the Paul Wild Observatory, Narrabri, New South Wales, Australia. The ATCA consists of 6$\times$22 metre antennas, 5 of which lie on a 3 km east-west railway track with the sixth antenna located 3 km farther west. Each antenna is fitted with a dual feedhorn system allowing simultaneous measurements of two wavelengths, either 20 \& 13 cm or 6 \& 3.6 cm. Additionally, during a recent upgrade, 12 and 3 mm receivers were installed allowing observations at $\sim$ 20 and $\sim$ 95 GHz respectively. The 6/3.6 cm receiver system was used for the observations of SFO~58, SFO~68 and SFO~76, while observations of SFO~75 were carried out using the new 12 mm receivers. A summary of the observational parameters is presented  in Table~\ref{tbl:radio_observations}.

\begin{table*}
\begin{center}
\caption[ATCA radio observational parameters]{Observational parameters for the ATCA radio observations.}
\begin{tabular}{ccccccc}
\hline
\hline
Cloud& Wavelength&\multicolumn{2}{c}{Phase centre} & Array & Integration& Phase\\
 id.&(cm)& $\alpha$(J2000) & $\delta$(J2000)& configurations& time (hrs) & calibrators\\
\hline
SFO~58\dotfill&6/3.6& 08:45:25.4 & $-$41:16:02& 750/352/214 &36& 0826--373\\
SFO~68\dotfill&6/3.6& 11:35:31.9 & $-$63:14:51& 750/352/367&18&  1129--58\\
SFO~76\dotfill&6/3.6& 16:10:38.6 & $-$49:05:52& 750/352/367&18& 1613--586  \\
\hline
SFO~75\dotfill&1.3&  15:55:50.4 & $-$54:38:58& 352&12& 1613-586  \\
\hline
\label{tbl:radio_observations}
\end{tabular}
\end{center}
\end{table*}

The 6/3.6 cm observations of SFO~58, SFO~68 and SFO~76 were made at two different frequency bands centred at 4800 and 8309~MHz using bandwidths of 128 and 8~MHz respectively. The second frequency was observed in spectral band mode (using a total of 512 channels, giving a frequency resolution of 15.6~kHz) in order to observe the H92$\alpha$ radio recombination line, however, this proved too faint to be detected toward all three clouds. Each source was observed using three separate configurations over a twelve hour period for SFO~58, and six hours each for SFO~68 and SFO~76 (split into 8$\times$40 minute observations spread over a 12~hour period to optimise \emph{u-v} coverage). 

Observations of SFO~75 were made using the 12 mm receivers centred at 23569 MHz ($\sim1.3$~cm) and used a bandwidth of 128 MHz. SFO~75 was observed for a total of 12 hours in a single configuration.

To correct these data for fluctuations in the phase and amplitude caused by atmospheric and instrumental effects, a phase calibrator was observed for two minutes after approximately every 40 minutes of on-source integration ($\sim$ 15 minutes for SFO~75 due to the atmosphere being less stable at higher frequencies). The primary flux calibrator, 1934--638, was observed once during each set of observations to allow for the absolute calibration of the flux density. To calibrate the bandpass the bright point source 1921--293 was also observed once during each set of observations. The phase centres, array configurations, total integration time and phase calibrators are tabulated in Table~\ref{tbl:radio_observations}.

The calibration and reduction of these data were performed using the MIRIAD reduction package
(\citealt{sault1995}) following standard ATCA procedures. The data were CLEANed using a robust weighting of 0.5 to obtain the same sensitivity as natural weighting, but with a much improved beam-shape and lower sidelobe contamination, with the exception of the 3.6 cm data for SFO~58 and SFO~68. Using a robust weighting of 0.5 for these two clouds resulted in poorer imaging of the large scale structure of the ionised gas surrounding their bright rims. (The smaller bandwidth used for the 3.6 cm observations results in these images being a factor of three less sensitive than the 6 cm observations.) For the 3.6 cm data for SFO~58 and SFO~68 a robust weighting of 1 was used, resulting in a slight loss of resolution but improved sensitivity. The data obtained from baselines which included the 6th antenna were found to distort the processed images (due to the large gap in \emph{u-v} coverage at intermediate baselines) and so were excluded from the final images. There are three calibration errors that need to be considered: absolute flux calibration, r.m.s pointing errors and the lack of short baselines. The uncertainties introduced by the first two of these are no more than a few percent each. The shortest baseline for all of these radio observations was 31 metres, and therefore flux loss due to short baselines is not considered to be significant. The combined uncertainty is estimated to be no more than $\sim$ 10 \%.

\section{Results and analysis}

\begin{figure*}[!]
\begin{center}
\includegraphics[width=0.45\textwidth,trim=10 0 10 10]{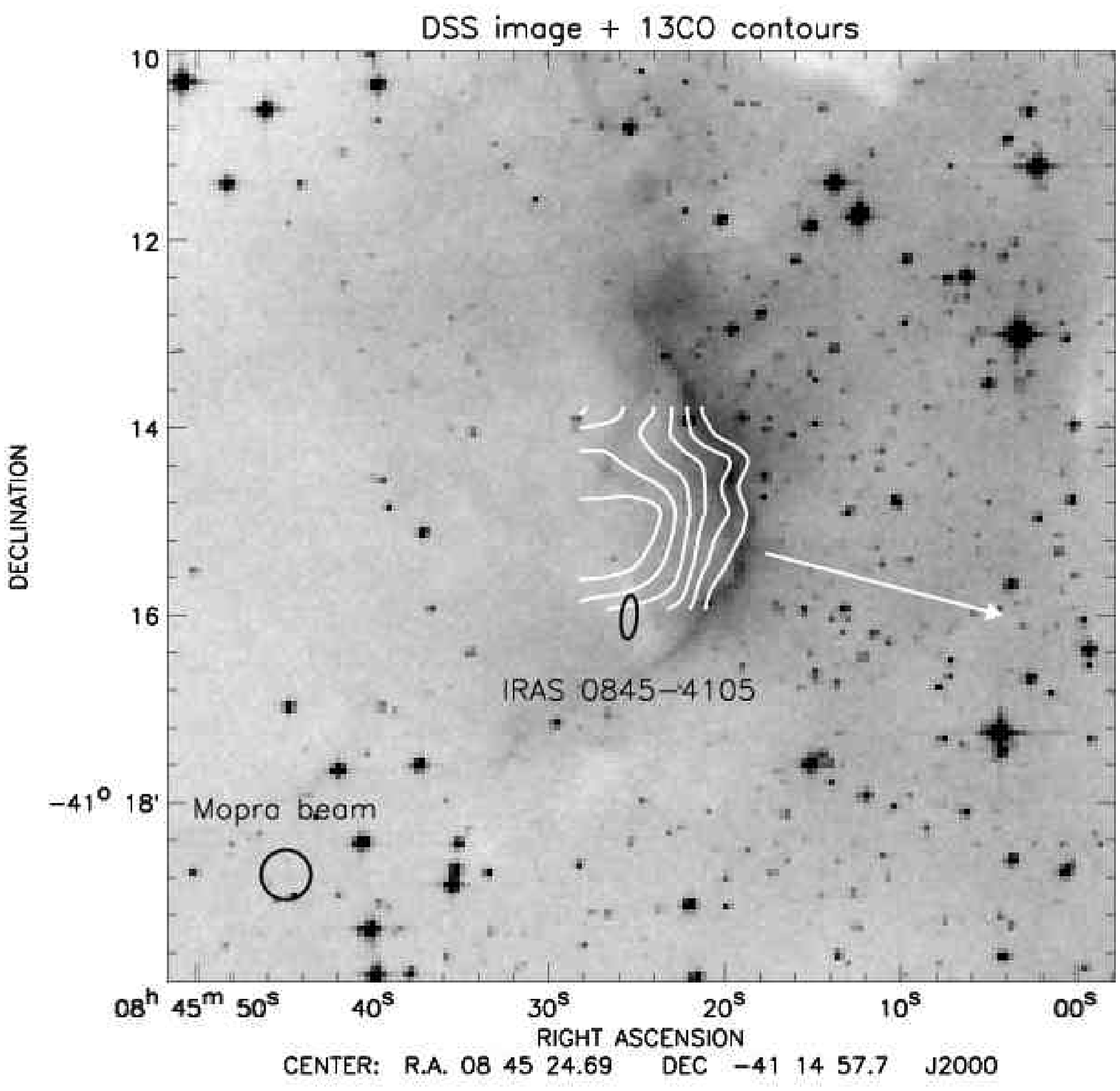}
\includegraphics[width=0.45\textwidth,trim=10 0 10 10]{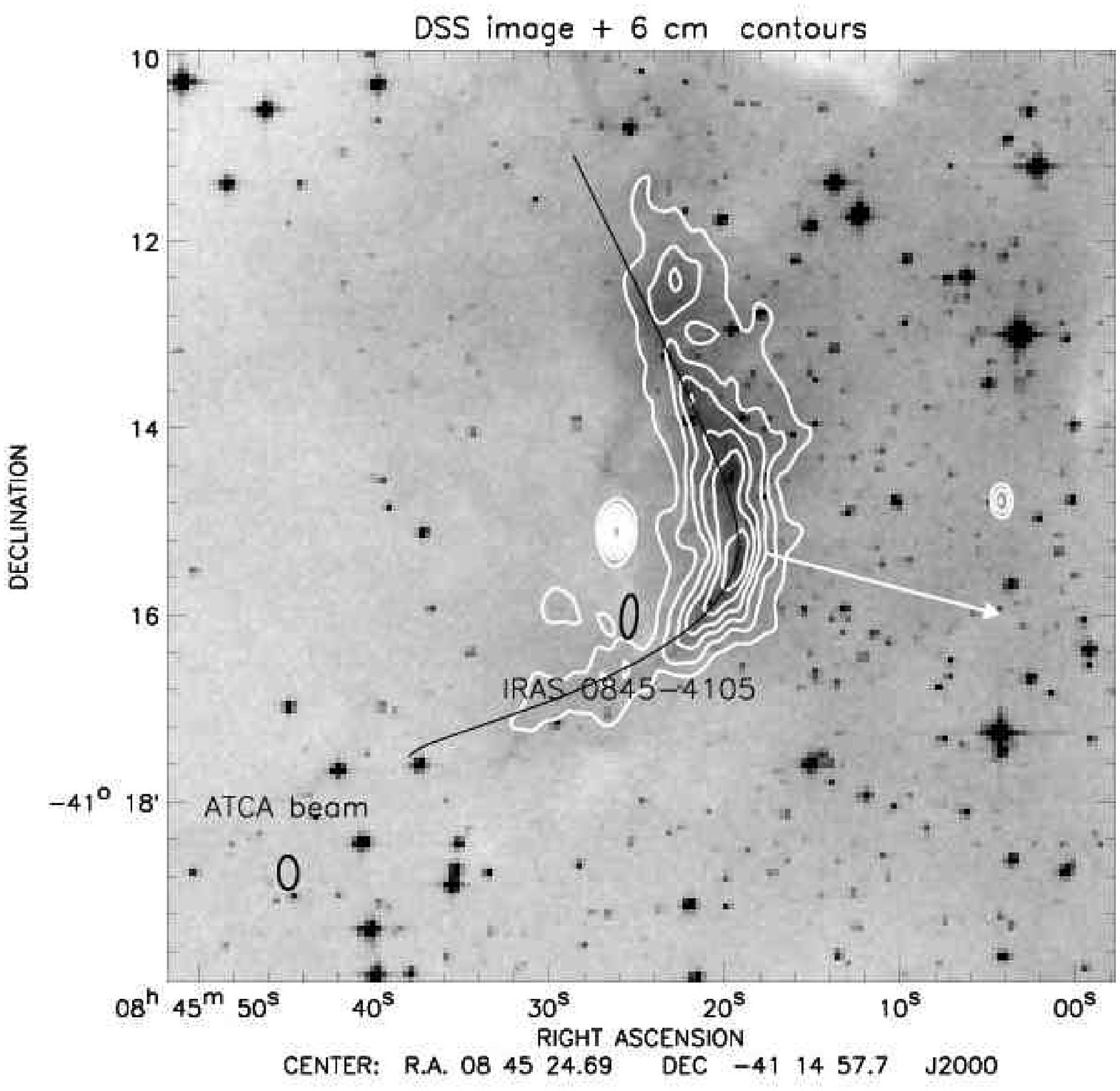}\\
\includegraphics[width=0.43\textwidth, height=0.45\textwidth, trim=0 0 -50 0]{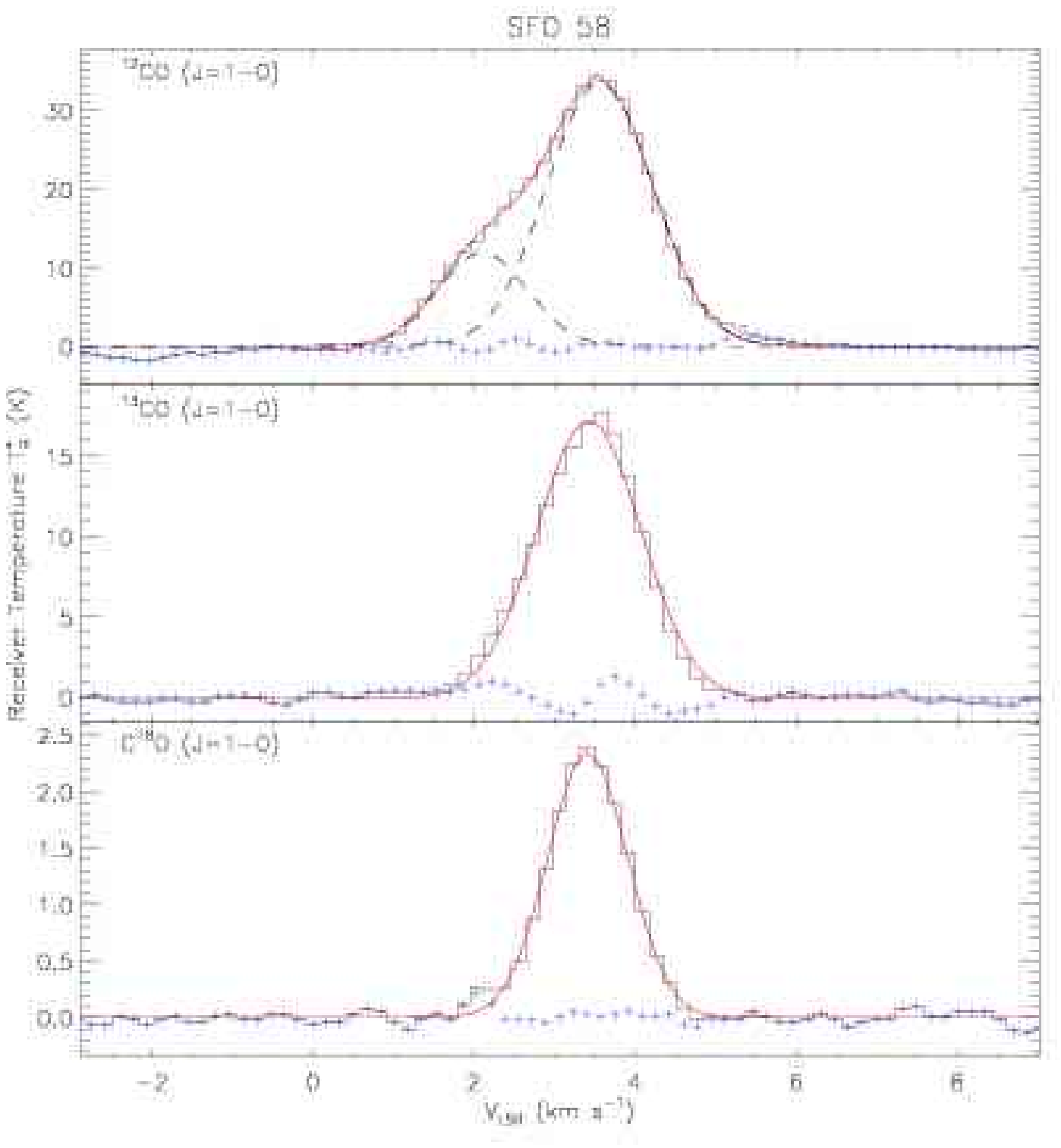}
\caption[Results of CO observations toward SFO~58]{DSS (R band) image of SFO~58 overlaid with contours of the integrated $^{13}$CO emission (\emph{upper left panel}), 6 cm radio continuum emission (\emph{upper right panel}) and source-averaged $^{12}$CO, $^{13}$CO and C$^{18}$O spectra (\emph{lower panel}). The $^{13}$CO contours begin at 30 \% of the peak emission and increase in steps of 10 \% of the peak emission (i.e. 29.1 K \kms). Radio contours start at 6$\sigma$ (1$\sigma\sim  0.10$ mJy) and increase in steps of 6$\sigma$. The position of the associated IRAS point source and its positional uncertainty is indicated by a black ellipse. The arrow indicates the direction of the candidate ionising star identified in Table \ref{tbl:HII_regions}. The Mopra and ATCA beam sizes are shown in the bottom left corner of their respective contoured images. A black curved line has been fitted to the contours of minimum slope of the radio contoured image (\emph{upper right panel}) in order to emphasis the rim morphology (see Section~5.3). \emph{Lower  panel}: Core-averaged spectra of the three observed molecular transitions are plotted in the form of a histogram, the Gaussian fits to these data and the corresponding residuals are indicated by a continuous (red) line and (blue) crosses respectively. These spectra were obtained by averaging all spectra within the FWHM of the molecular core (see text for details).}
\label{fig:co_sfo58}
\end{center}
\end{figure*}

\begin{figure*}[!htb]
\begin{center}
\includegraphics[width=0.45\textwidth,trim=10 0 10 10]{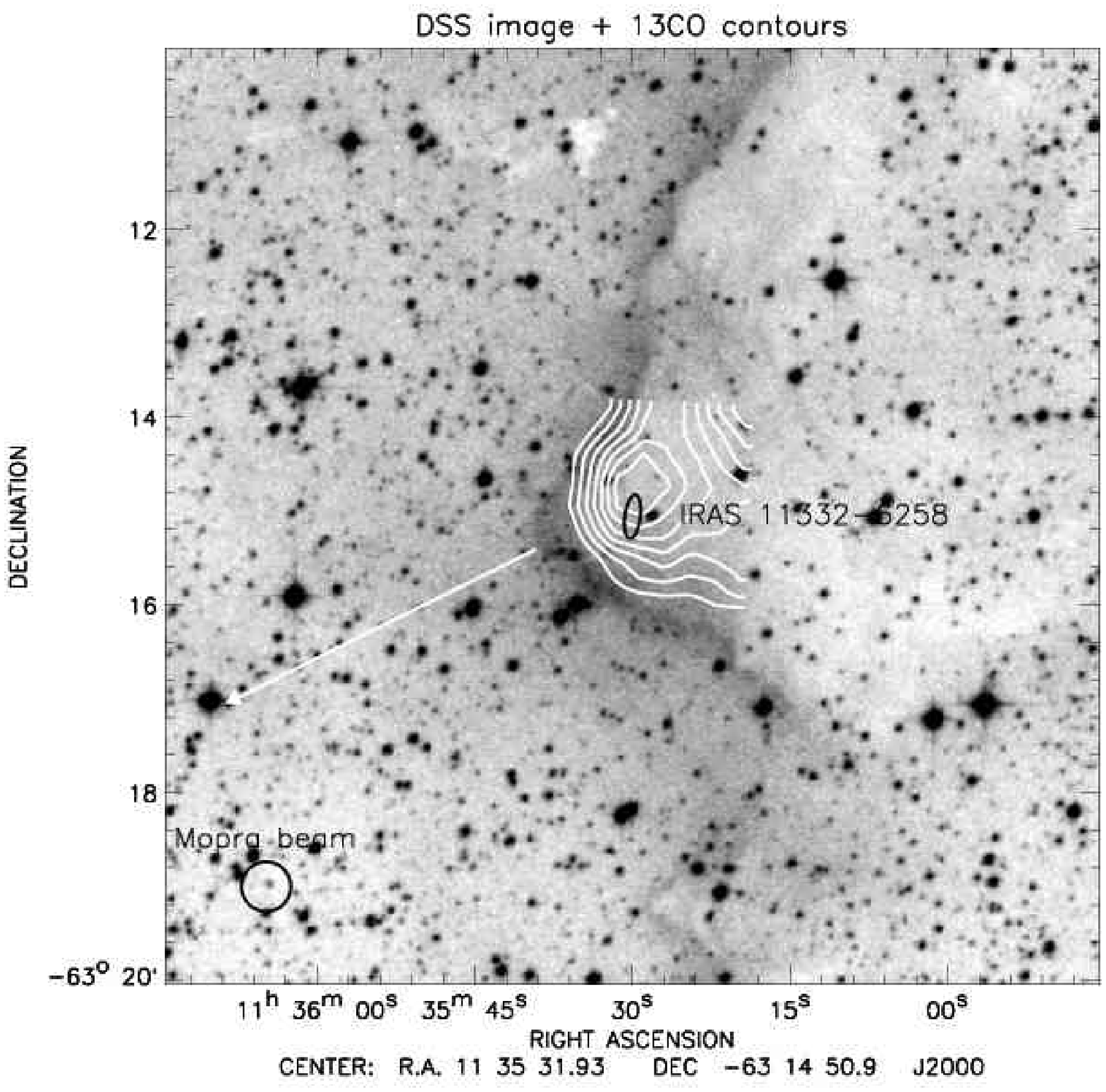}
\includegraphics[width=0.45\textwidth,trim=10 0 10 10]{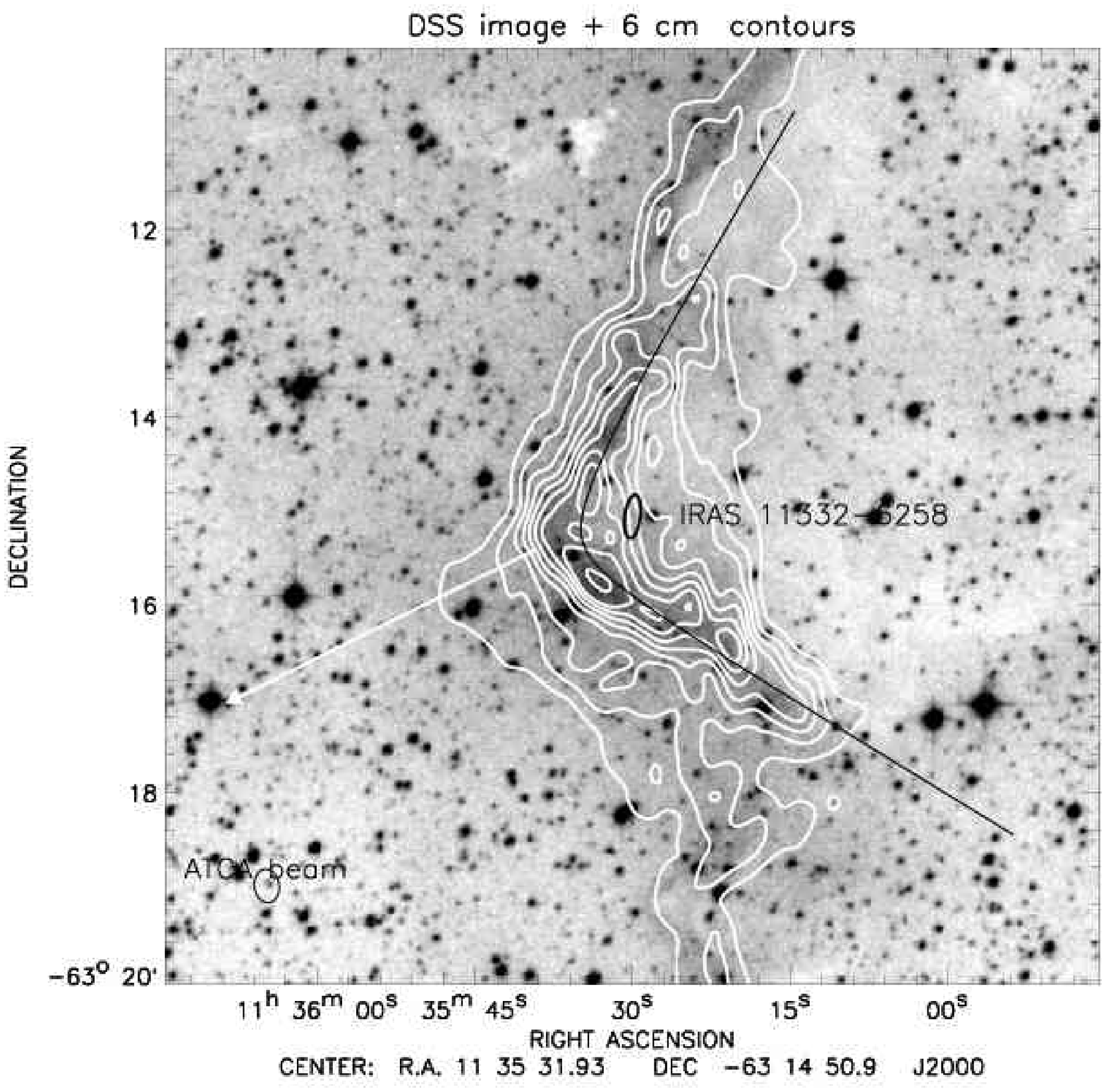}\\
\includegraphics[width=0.45\textwidth,trim=10 0 10 10]{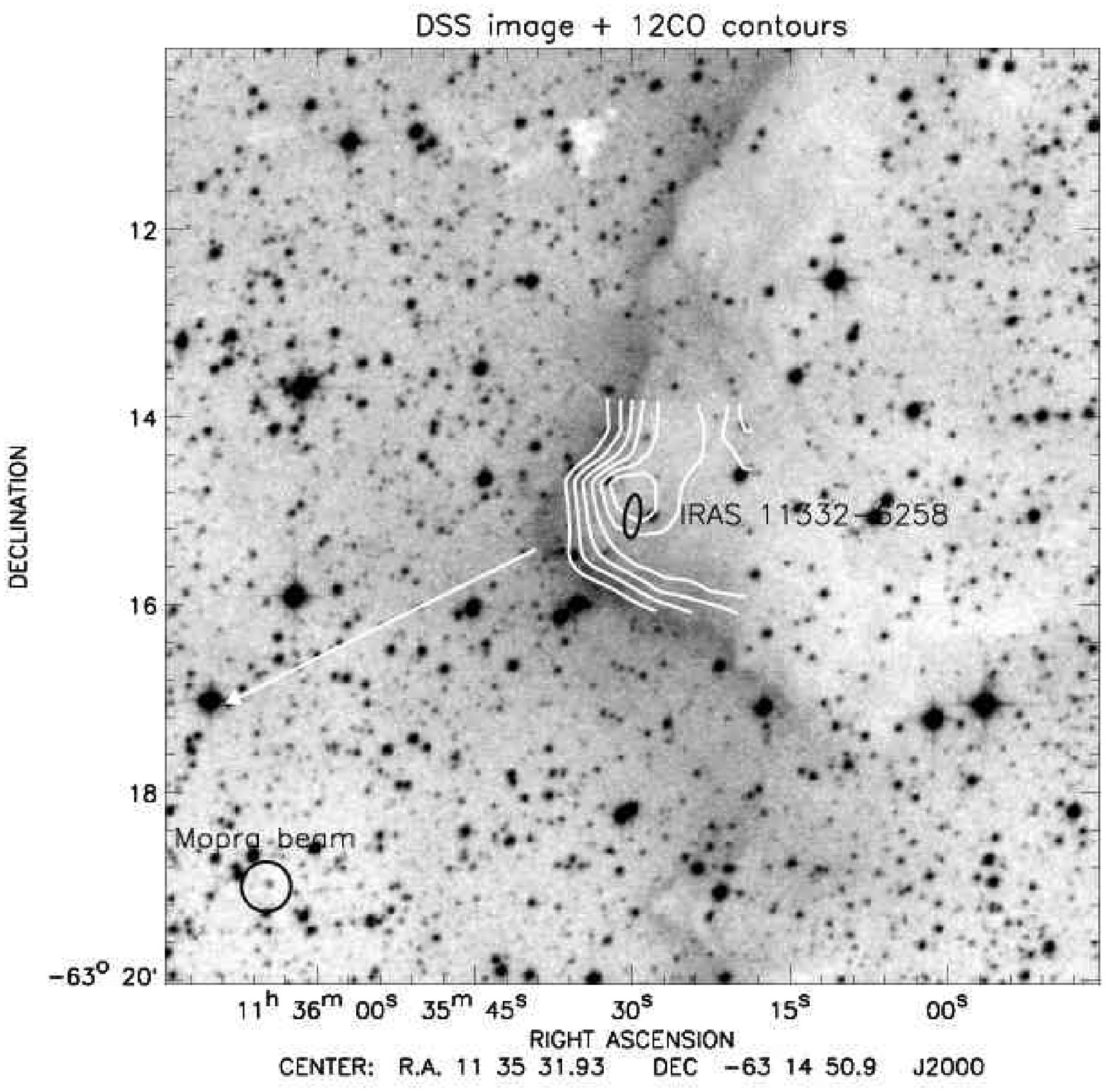}
\includegraphics[width=0.43\textwidth,height=0.45\textwidth,trim=0 -50 0 0]{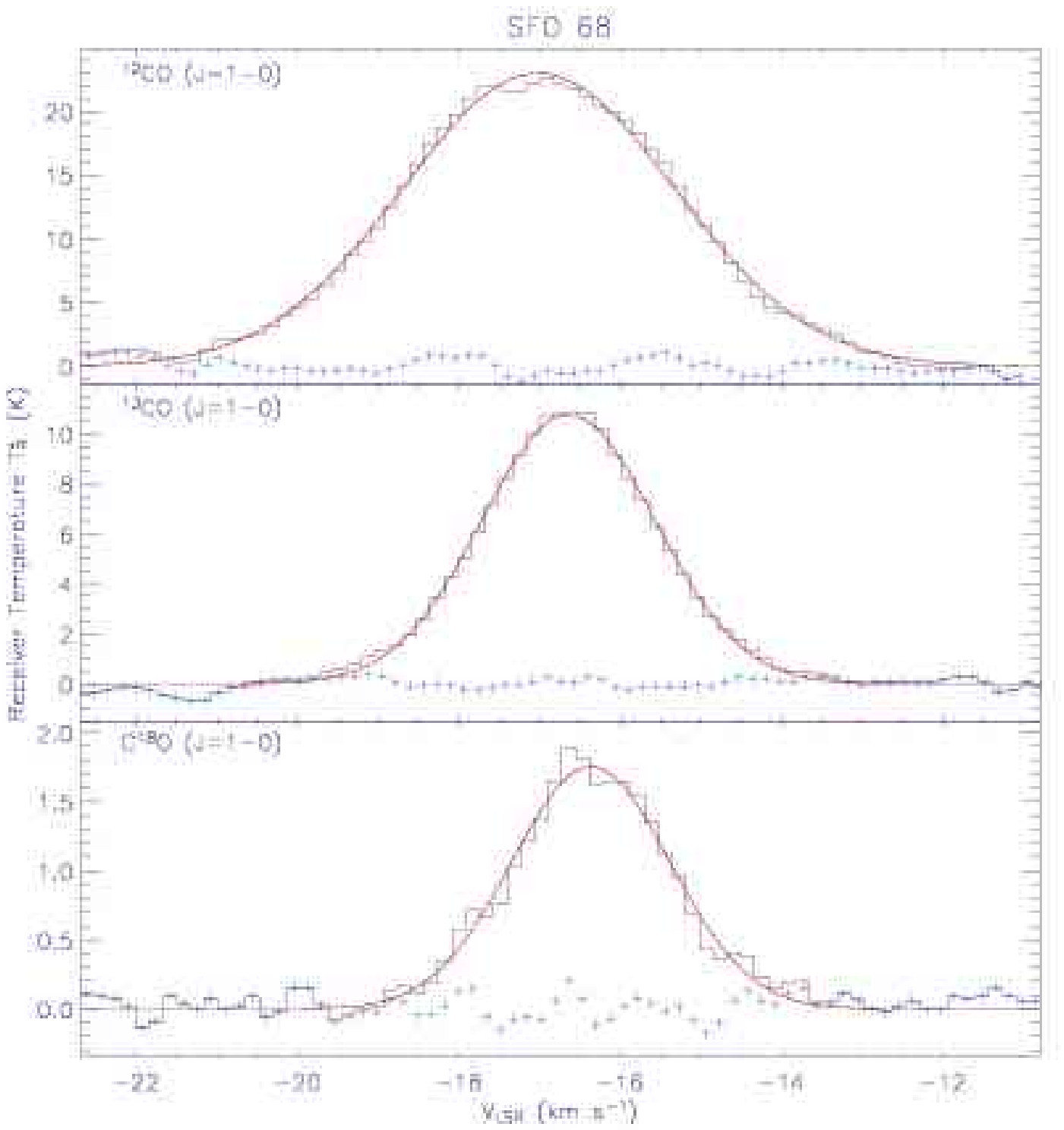}\\

\caption[Results of CO observations toward SFO~68]{SFO~68: As Figure~\ref{fig:co_sfo58} except integrated $^{12}$CO emission (\emph{lower left panel}) and source-averaged $^{12}$CO, $^{13}$CO and C$^{18}$O spectra (\emph{lower right panel}). The CO contours begin at 30 \% of the peak emission and increase in steps of 10 \% of the peak emission (i.e. 35.2 K \kms~and 97.9 K \kms~for the $^{13}$CO and $^{12}$CO maps respectively). Radio contours start at 3$\sigma$ (1$\sigma\sim  0.16$ mJy) and increase in steps of 3$\sigma$.} 
\label{fig:co_sfo68}

\end{center}
\end{figure*}

\begin{figure*}[!htb] 
\begin{center}
\includegraphics[width=.45\textwidth,trim=10 0 10 0]{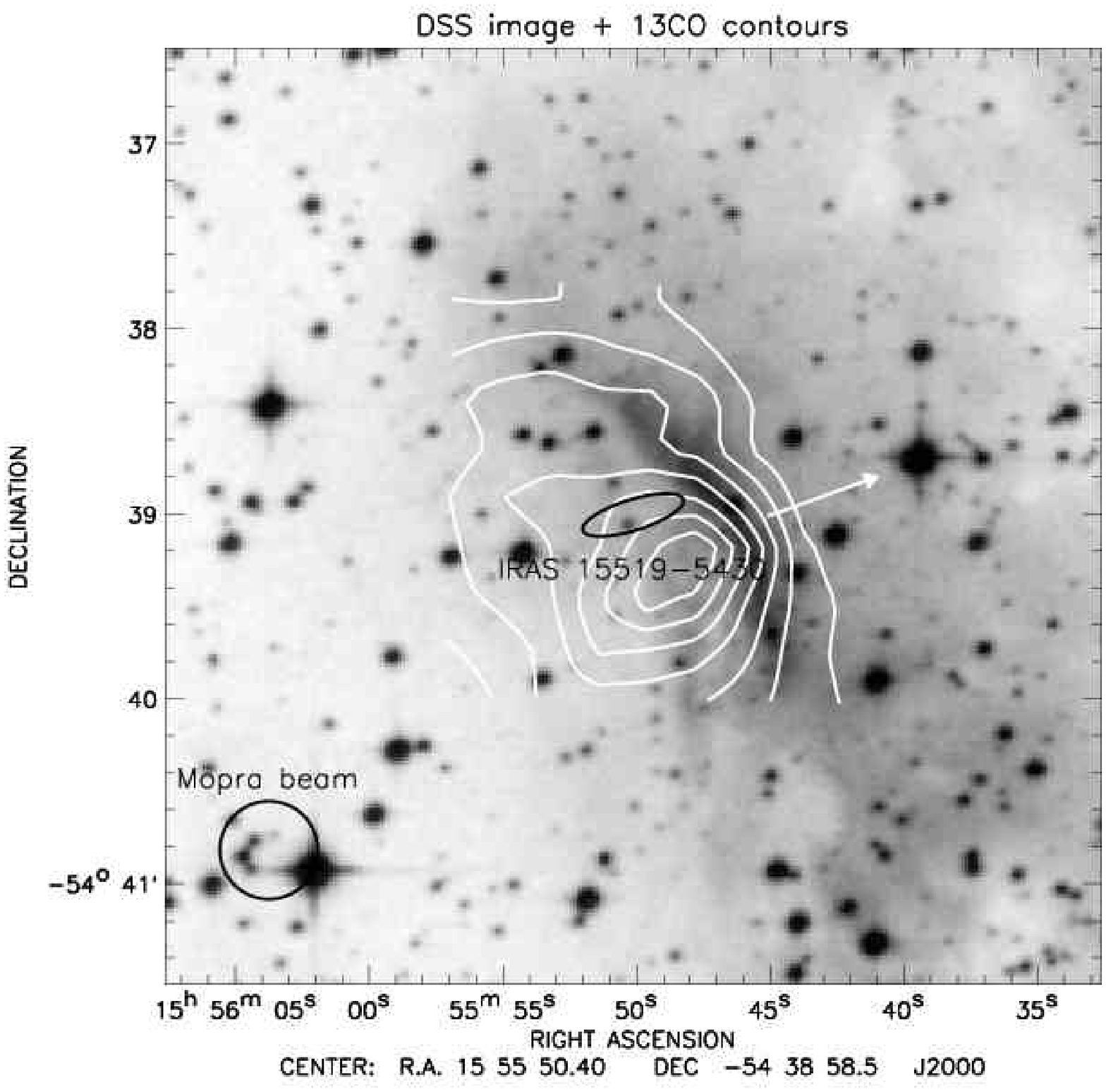}
\includegraphics[width=.45\textwidth,trim=10 0 10 0]{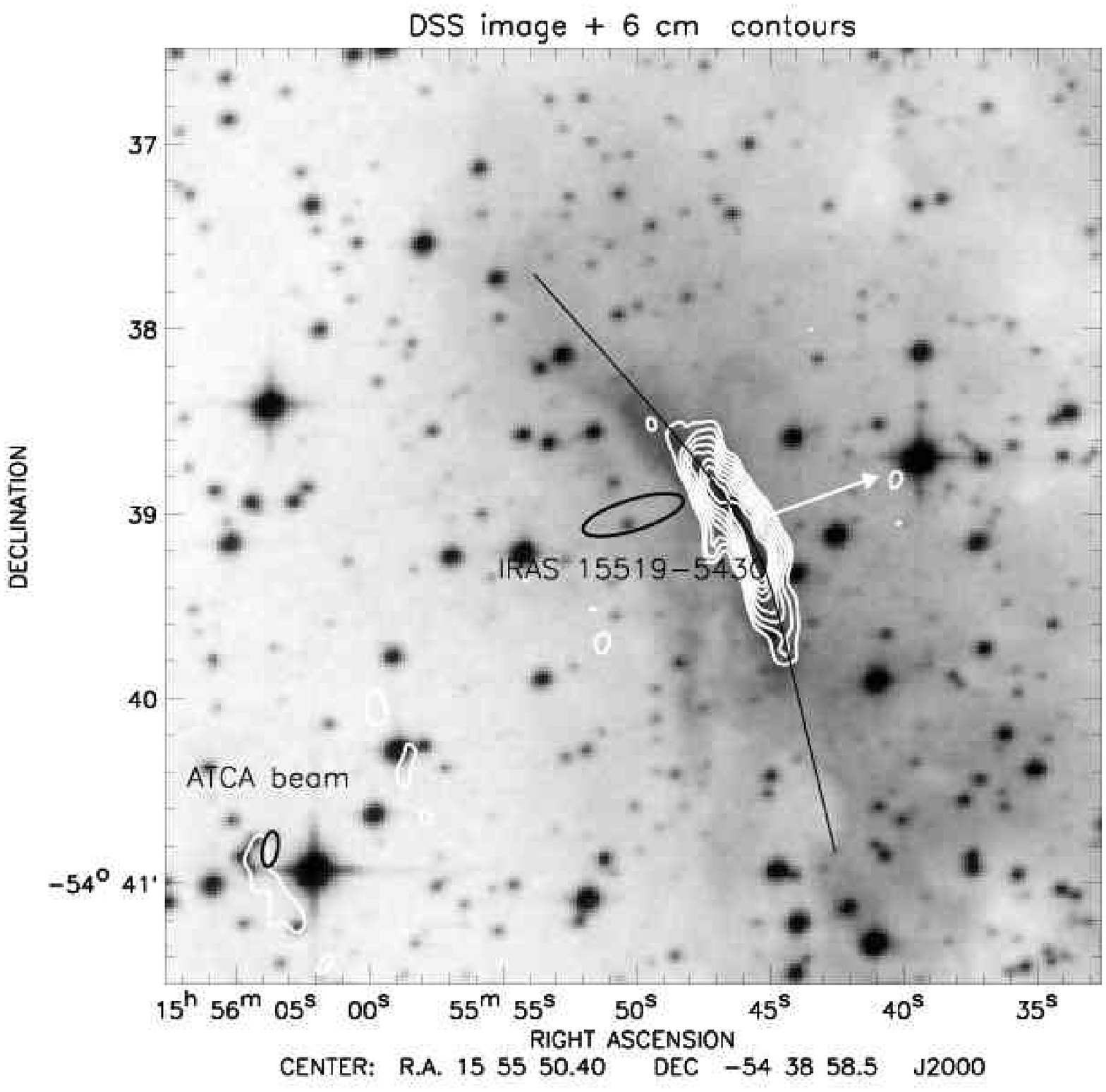}\\
\includegraphics[width=.45\textwidth,trim=10 0 10 0]{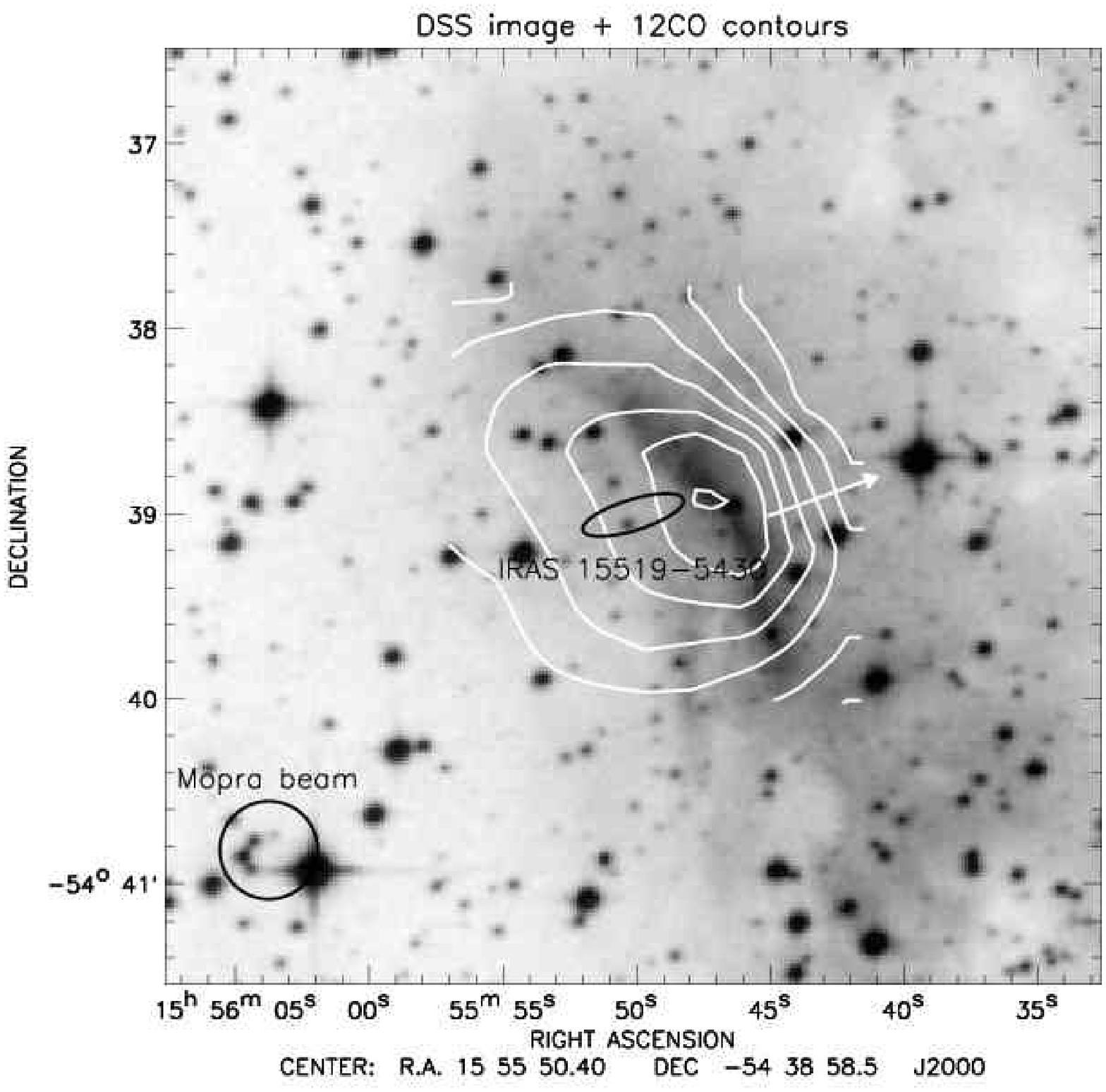}
\includegraphics[width=0.43\textwidth,height=0.45\textwidth,trim=0 -50 0 0]{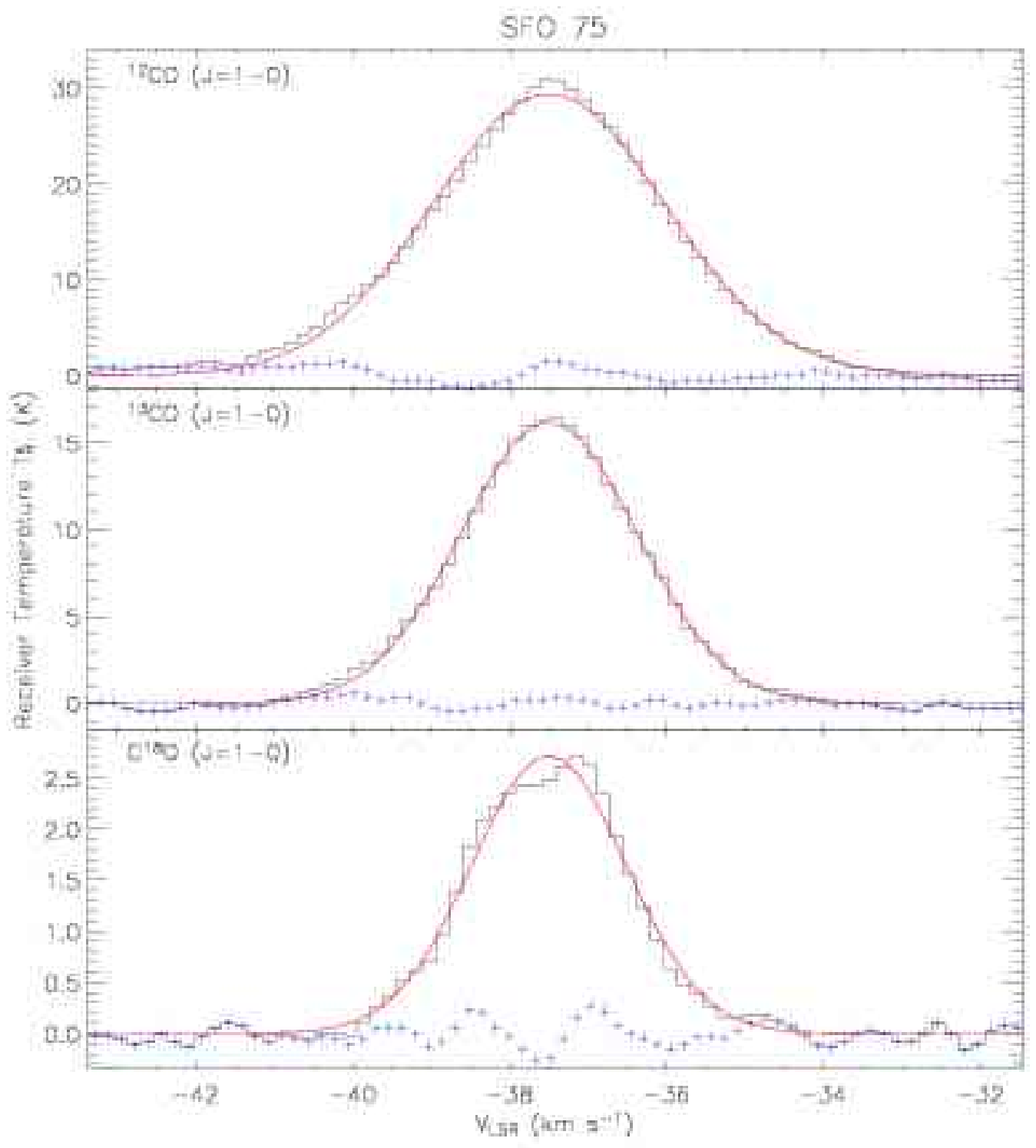}\\
\caption[Results of CO observations toward SFO~75 and SFO~76]{SFO~75: As for Figure~\ref{fig:co_sfo68}. The CO contours begin at 30 \% of the peak emission and increase in steps of 10 \% of the peak emission (i.e. 47.7 K \kms~and 146.5 K \kms~for the $^{13}$CO and $^{12}$CO maps respectively). Radio continuum emission is at 1.3 cm with contours starting at 5$\sigma$ (1$\sigma\sim$  0.20 mJy) and increase in steps of 2$\sigma$.}
\label{fig:co_sfo75}
\end{center}
\end{figure*}

\begin{figure*}[!htb] 
\begin{center}
\includegraphics[width=.45\textwidth,trim=10 0 10 0]{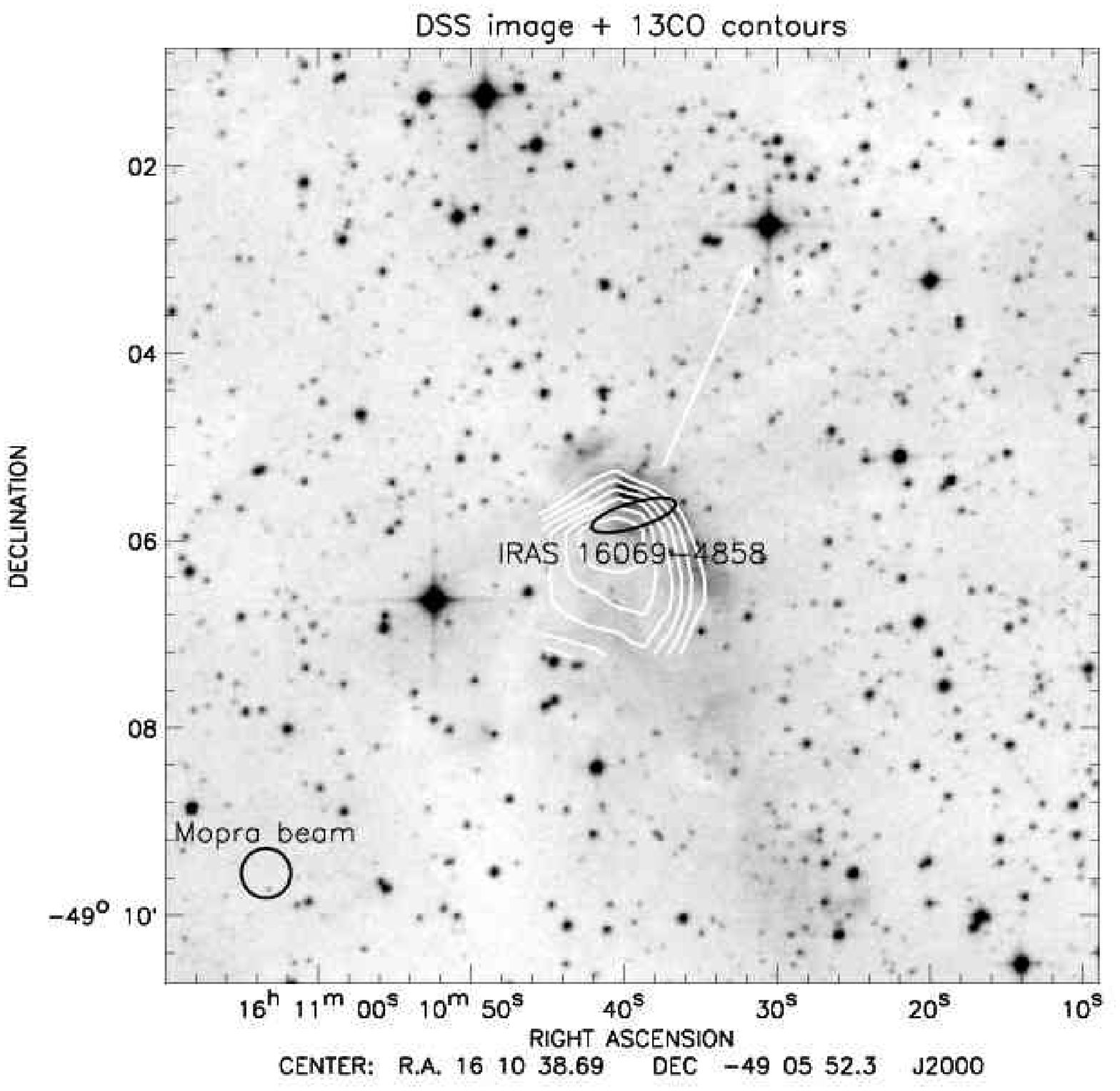}
\includegraphics[width=.45\textwidth,trim=10 0 10 0]{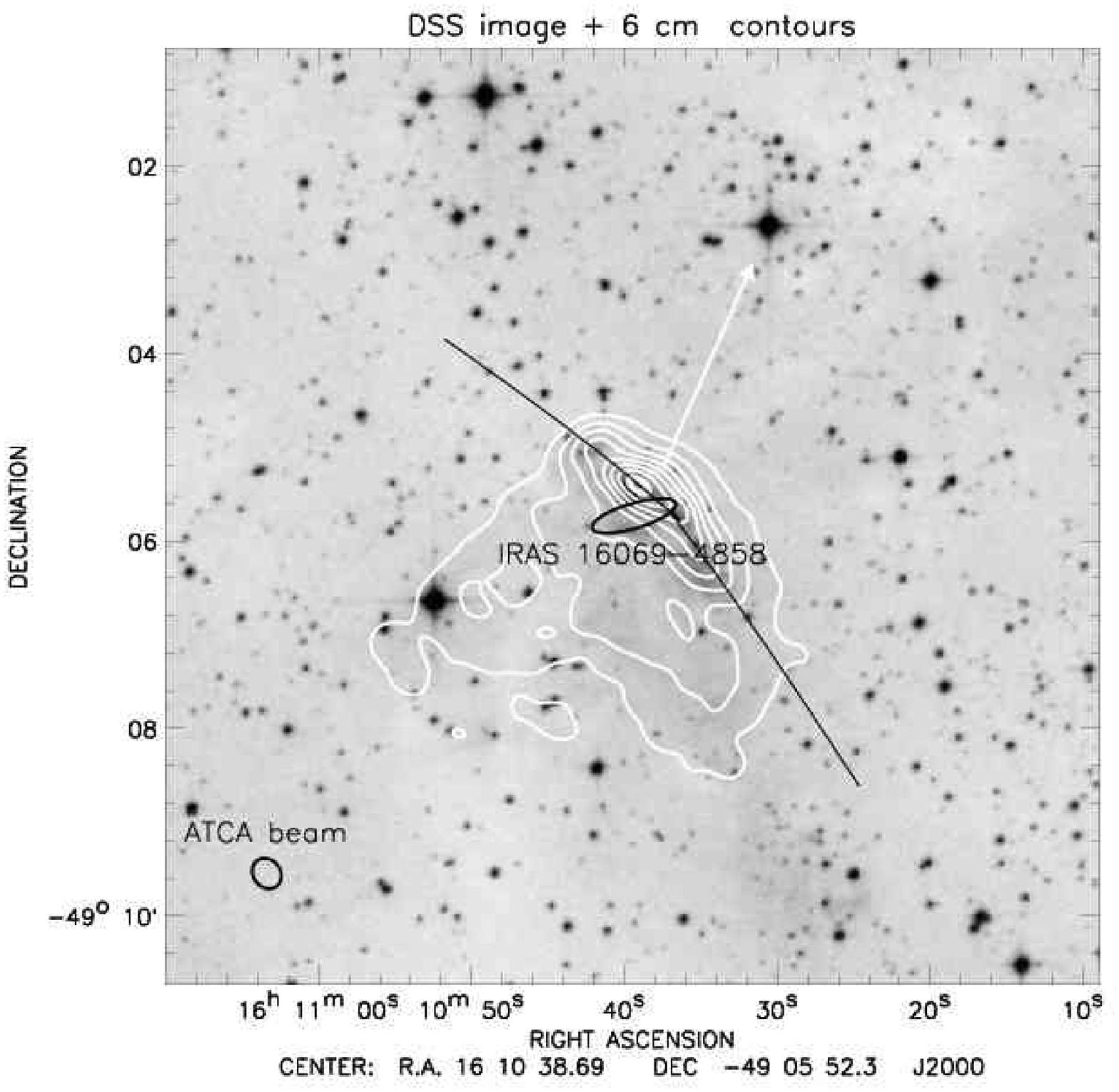}\\
\includegraphics[width=.45\textwidth,trim=10 0 10 0]{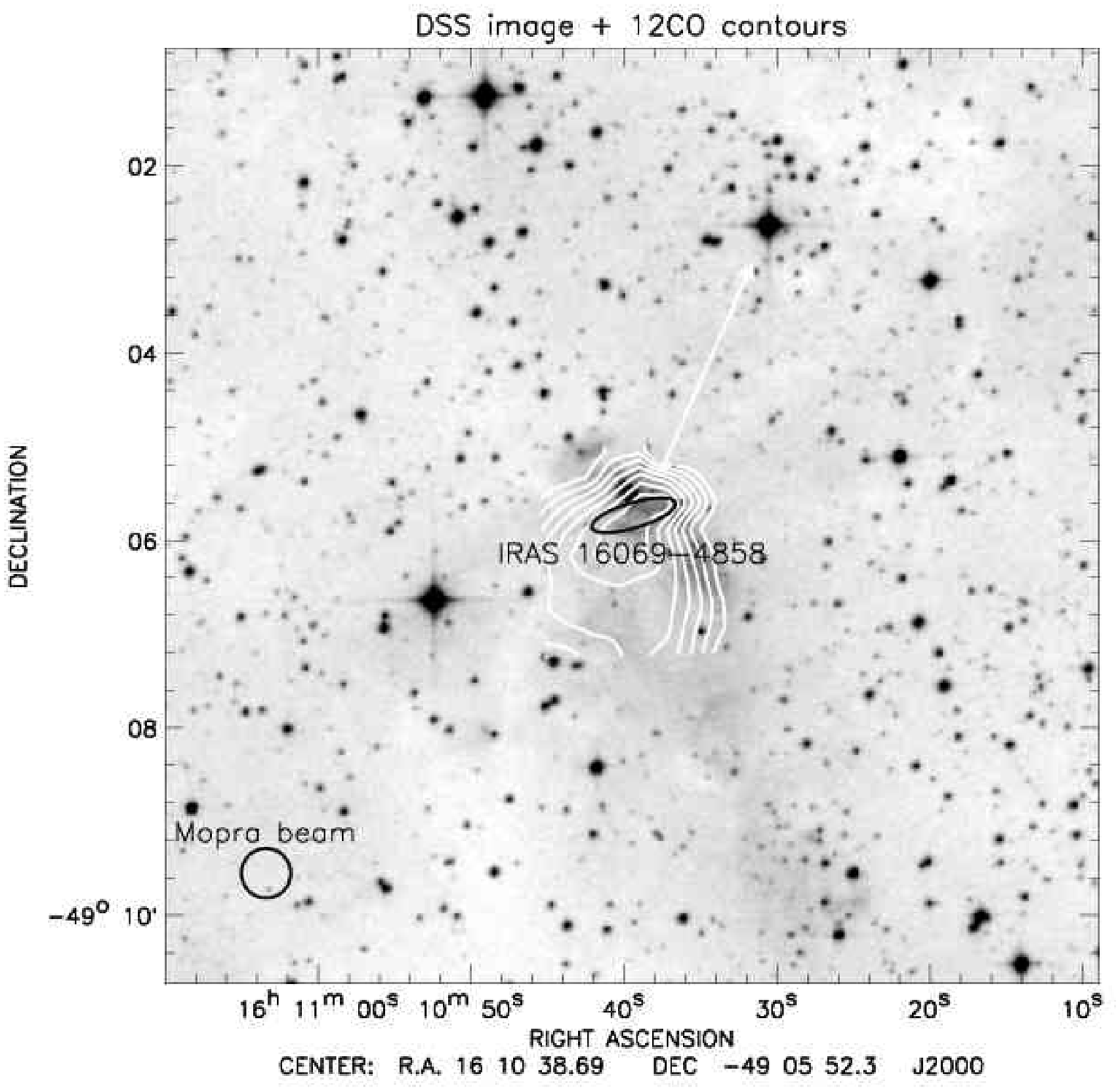}
\includegraphics[width=0.43\textwidth,height=0.45\textwidth,trim=0 -50 0 0]{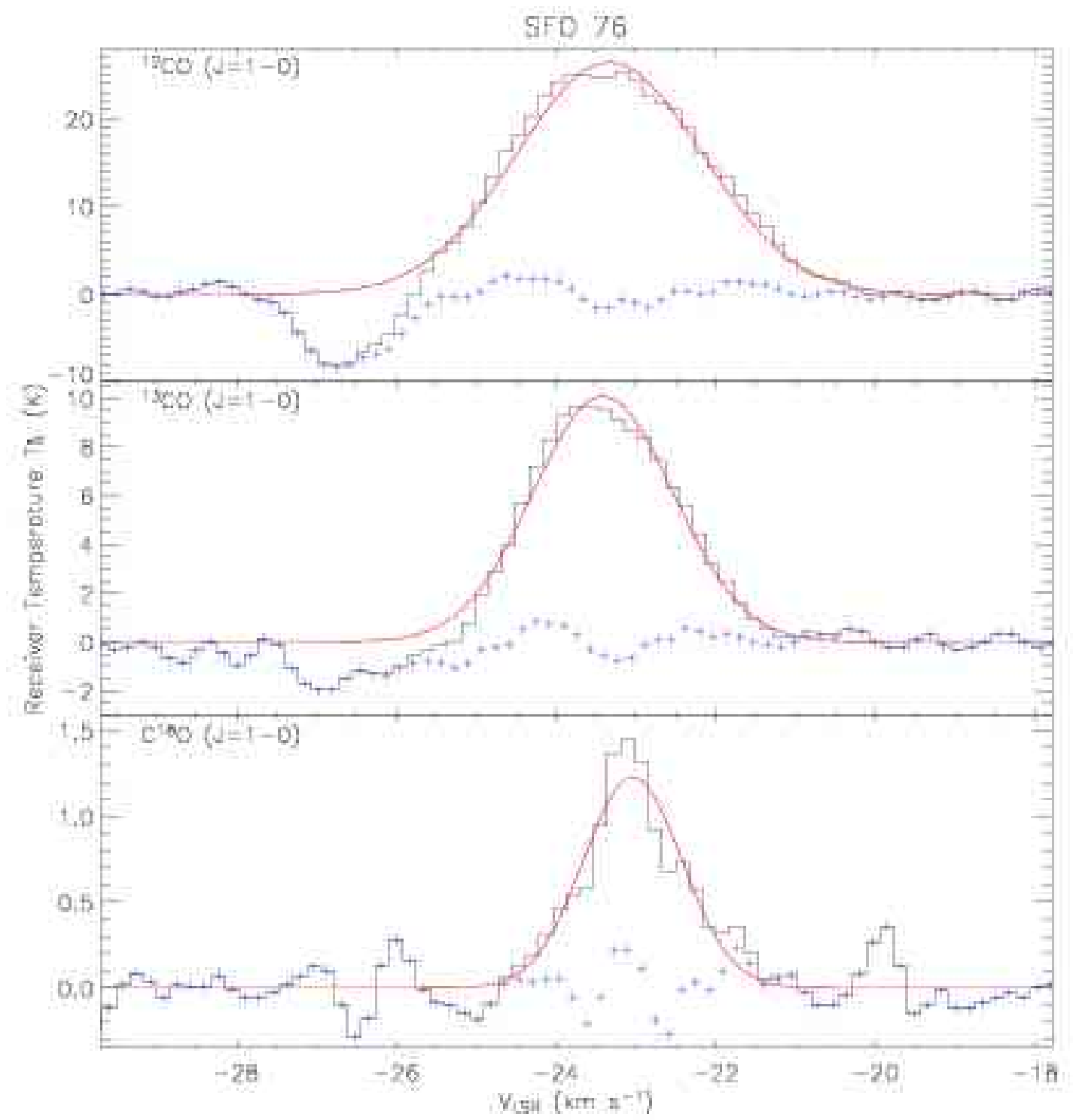}\\
\caption[Results of CO observations toward SFO~75 and SFO~76]{SFO~76: As for Figure~\ref{fig:co_sfo68}. The CO contours begin at 30~\% of the peak emission and increase in steps of 10~\% of the peak emission (i.e. 22.8 K \kms\ and 173.6~K~\kms\ for the $^{13}$CO and $^{12}$CO maps respectively). Radio contours start at 6$\sigma$ (1$\sigma\sim  0.30$~mJy) and increase in steps of 6$\sigma$.}
\label{fig:co_sfo76}
\end{center}
\end{figure*}

In Figures \ref{fig:co_sfo58}--\ref{fig:co_sfo76} we present plots of the integrated $^{13}$CO (\emph{upper left
panel}), $^{12}$CO (\emph{lower left panel}), radio continuum emission (\emph{upper right panel}) contoured over a
DSS image of the BRC and surrounding region. These images reveal a strong spatial correlation between the
distribution of both the molecular and ionised gas with the optical morphology of the bright rims. 

The molecular gas traced by the $^{12}$CO and $^{13}$CO contours displays a similar morphology, both tightly
correlated within the rim of the cloud and with a steep intensity gradient decreasing toward the HII region.
The steep intensity gradient is possibly a consequence of shock compression of the molecular gas which is swept up
in front of the expanding ionisation front and accelerated into the cloud, consistent with the predictions of RDI.
The $^{13}$CO emission reveals the presence of a dense molecular core embedded within every BRC, set back slightly
from the bright rim of the cloud with respect to the direction of the ionising star(s). All of the molecular cores
appear to be centrally condensed, which suggests they may be gravitationally bound, or have a gravitationally bound
object embedded within them, such as a  protostar. 

There is an interesting difference between the spatial distribution of $^{12}$CO and $^{13}$CO emission toward SFO~75. The peak of the integrated $^{12}$CO emission detected toward SFO~75 is correlated with the position of the bright rim of the cloud, and is slightly elongated in a direction parallel to the morphology of the rim, whereas the integrated $^{13}$CO emission peaks farther back within the molecular cloud and is elongated in a direction perpendicular to the rim. There are a few possible explanations that should be considered:  heating of the surface layers of the cloud by the FUV radiation field, the ionisation front is preceeded by a PDR, which could sharpen the $^{12}$CO emission, or that the cloud is angled to the line of sight and that we are seeing the body of the cloud through the bright rim.

The morphology of the radio and optical emission seen toward all four clouds shows the rims to be curving directly away from the ionising star(s), starting to form the typical cometary structure seen toward more evolved BRCs (e.g. the Eagle nebula). The presence of radio emission and its tight correlation with the morphology of the rim strongly supports the hypothesis that an IBL is present between the ionising star(s) and the molecular material within these BRCs. In addition to the IBL, the radio emission image of SFO~58, also reveals the presence of a compact radio source within the optical boundary of the cloud, which is coincident with the position of the molecular core identified in $^{13}$CO image, possibly indicating the presence of an UC HII region within this cloud. 

The radio images were analysed using the visualisation package \emph{kvis} (part of the \emph{karma} image analysis
suite (\citealt{gooch1996}). The image parameters and measurements of the peak and source integrated emission are
summarised in Table~\ref{tbl:image5}.

\begin{table*}
\caption[Summary of physical parameters derived from radio observations]{Summary of physical parameters derived from radio observation images.}
\begin{center}
\small
\begin{tabular}{cccccccc}
\hline
\hline
Cloud  &  &  Restoring &Position &Peak   & Integrated & Source-averaged &  Image  \\
id.& $\lambda$  &    beam &angle & emission & emission & emission&  r.m.s. noise\\
&	(cm)& (arcsec)& (degrees)&(mJy/beam) & (mJy) & (mJy/beam)&(mJy/beam) \\
\hline
SFO~58 & 3.6  &$27.6\times17.6$ & 2.8  &2.66 &23.1& 1.33  &   0.11\\
& 6  &  $21.6\times12.6$ & 1.7 &   2.38 &38.5& 1.44    &  0.10 \\
\hline
SFO~68	& 3.6 &$20.9\times15.6$ & 7.5 & 5.22 & 69.3 & 1.63 & 0.17 \\
& 6  & $21.6\times15.6$ & 10.42 & 6.36  & 151.1 & 2.97 & 0.16\\
\hline
SFO~76 & 3.6  & $13.3\times10.9$ & 50.1  & 9.84 & 195.4 & 1.74 & 0.25\\
& 6 & $20.6\times17.3$ & 35.4 & 20.82 & 255.7 & 6.99 &  0.30\\
\hline
SFO~75   & 1.3  & $11.5\times4.9$ & $-$9.2 & 3.39 & 30.3 & 0.53 & 0.20 \\
\hline
\end{tabular}
\label{tbl:image5}
\end{center}
\end{table*}

The IRAS point sources within SFO~68, SFO~75 and SFO~76 are located slightly behind the ionisation front, with respect to the ionising star(s), and toward the centre of the bright rim, where one would expect photoionisation induced shocks to focus molecular material, and where the RDI models predict the cores to form (\citealt{lefloch1994}). The positions of the $^{13}$CO peak emission and the IRAS point source seen toward SFO~58 (Figure~\ref{fig:co_sfo58}) are not so well correlated; the IRAS point source is located approximately 1\arcmin~to the south of the molecular peak. It is possible that the IRAS point source is unrelated to the molecular core detected, but is associated with another core which has formed on the edge of the molecular cloud, however, considering the positional correlation of the $^{13}$CO core with the compact radio source and taking account of the  pointing errors, size of the IRAS beam ($\sim2^{\prime}$ at 100 $\mu$m), and the positional inaccuracy of the IRAS point source, we do not consider the displacement between the IRAS point source and the centre of the $^{13}$CO core to be significant.\footnote{Comparing the positions of dense cores identified from ammonia maps of the Orion and Cepheus clouds with the positions of their associated IRAS sources \citet{harju1993} found that they could be offset from each other by up to 80$^{\prime\prime}$.}

\subsection{Physical parameters of the cores}
\label{sect:co_analysis}
The angular size of each of the four molecular cores was estimated from the FWHM of azimuthally averaged $^{13}$CO intensity maps. Averaging all spectra within the derived angular size of each core, a source-averaged spectrum was produced for each of the three molecular lines; these spectra are presented in the \emph{lower right panels} of Figures~\ref{fig:co_sfo58}--\ref{fig:co_sfo76}. We note that the baseline for SFO~76 is poor due to the fact that  the reference position was found to contain emission at a nearby velocity (i.e. $-$26.5 km s$^{-1}$) to the main line.

Gaussian profiles were fitted to the core-averaged spectra of each core to determine the emission peak, FWHM line width and V$_{\rm{LSR}}$ for each spectral line. These values are listed in Table~\ref{tbl:co_data5} along with the central position of each core derived from a 2D Gaussian fit to the $^{13}$CO emission map. The measured \vlsr~of each source was compared to those reported by \citet{yamaguchi1999} and found to agree to better than 2~\kms, with the exception of SFO~76. For this source \citet{yamaguchi1999} reported a V$_{\rm{LSR}}$ of $\sim$ $-$37 \kms~compared to our measurement of $-$23~\kms.  The \vlsr~obtained from our CO observations compares well with that reported by \citet{bronfman1996} from CS observations toward SFO~76 (i.e. $-$22.2~\kms). It is therefore unclear why there is such a large disagreement between the \vlsr~reported by \citet{yamaguchi1999} and that reported in this survey for SFO~76.

Comparisons between the $^{12}$CO lines and other isotopomers for each core reveal no significant variations in the kinematic velocities of the emission peaks for either SFO~75 or SFO~76. However, inspection of the SFO~58 source-averaged $^{12}$CO spectrum shows evidence of a blue wing component not present in either the $^{13}$CO or C$^{18}$O spectra (this spectrum was best fitted by two Gaussian components). Comparing the source-averaged $^{12}$CO spectrum of SFO~68 to the $^{13}$CO and C$^{18}$O spectra reveals a small shift in velocity, which suggests the presence of a broad blue wing. There are several physical phenomenon that could give rise to these observed line wings such as: another cloud on the line of sight, a signature of shock compressions in the surface layers, or the presence of a protostellar outflow.

\begin{figure}
\begin{center}
\includegraphics[width=.4\textwidth,trim=30 0 30 0]{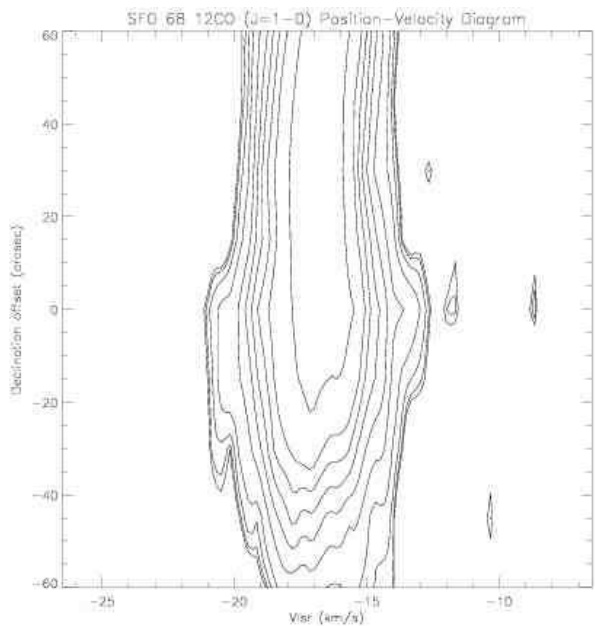}
\caption[Position-velocity diagram of SFO~68]{Position-velocity diagram of the $^{12}$CO emission observed toward SFO~68, taken through the peak of the molecular emission along the declination axes. Moderate velocity wing components can clearly be identified in this diagram.}
\label{fig:pv_sfo68}
\end{center}
\end{figure} 

In order to try to determine the source of these blue wings and to look for kinematic signatures of shocks we produced channel maps and position-velocity diagrams of the $^{12}$CO emission for each cloud. Only the position-velocity diagram produced for SFO~68 shows any evidence of a large scale velocity gradient, revealing the presence of moderate velocity wing components with a FWHM $\sim$ 8~\kms (see Figure \ref{fig:pv_sfo68}); these can either be attributed to a protostellar outflow, or could be tracing the compression/expansion motions of the surface layers of a collapsing cloud. The spatially localised wings seen in the position-velocity diagram are more suggestive of a bipolar outflow, however, our observations do not have either the mapping coverage or the angular resolution necessary to be able to spatially resolve the red and blue outflow lobes that would confirm the presence of an outflow.

The nature of the blue wings is at present unclear, however, taking into account the presence of a UC HII region within SFO~58 (see Section \ref{sect:uchii_region}) and the association of SFO~68 with methanol, OH and H$_2$O masers -- both of which are strong indications of ongoing star formation -- we consider the protostellar outflow hypothesis to be the most likely. We must stress that these are only tentative detections and further observations are required to confirm the presence of protostellar outflows in these two sources.
 
\begin{table*}
\caption[Physical values derived from CO spectra]{Results of Gaussian fitting of the CO spectra observed toward the molecular cores within BRCs.}
\begin{center}
\begin{tabular}{ccccccc}
\hline
\hline
Cloud id.&\multicolumn{2}{c}{Core position}&Molecular   & V$_{\rm{LSR}}$  & Peak T$_R^*$  &FWHM  \\
 & $\alpha$ (J2000) & $\delta$ (J2000) &line & (km s$^{-1}$) & (K) & (km s$^{-1}$)\\
\hline

SFO~58 &08:45:26 & $-$41:15:10 &$^{12}$CO& 2.1&  11.9& 1.3\\
&& && 3.6&  33.7& 1.51\\
&&&$^{13}$CO & 3.4&  17.3 & 1.5 \\
&&&C$^{18}$O & 3.4  &   2.4& 1.2 \\

\hline
SFO~68 & 11:35:31 & $-$63:14:31&$^{12}$CO& $-$17.1&  23.0& 4.0\\
&&&$^{13}$CO & $-$16.7 &   10.8& 2.5\\
&&&C$^{18}$O & $-$16.4 &   1.8&2.3\\
\hline
SFO~75 & 15:55:49 & $-$54:39:13&$^{12}$CO& $-$37.5  &  29.4& 3.5\\
&&&$^{13}$CO & $-$37.5  &  16.1& 2.6 \\
&&&C$^{18}$O & $-$37.5  &   2.7& 2.3\\
\hline
SFO~76 & 16:10:40 & $-$49:06:17&$^{12}$CO& $-$23.3  &  26.4& 2.7\\
&&&$^{13}$CO & $-$23.4 &  10.1& 2.1\\
&&&C$^{18}$O & $-$23.0 &   1.2& 1.5\\

\hline
\end{tabular}
\label{tbl:co_data5}
\end{center}
\end{table*}

The optically thin transitions of C$^{18}$O, and the moderately optically thick ($\tau$ $<$ 1) $^{13}$CO, were used
to determine the optical depth, gas excitation temperature and C$^{18}$O column density following the procedures
described by \citet{urquhart2004} using the following equations

\begin{equation}
\frac{T_{13}}{T_{18}}=\frac{1-{\rm{e}}^{-\tau_{13}}}{1-{\rm{e}}^{-\tau_{18}}}
\end{equation}

\noindent where $T_{13}$ and $T_{18}$ can be either the corrected antenna or receiver temperatures, and $\tau_{13}$ and $\tau_{18}$ are the optical depths of the $^{13}$CO and C$^{18}$O transitions respectively. The optical depths are related to each other by their abundance ratio 
such that $\tau_{13}$ = X$\tau_{18}$, where X is the $^{13}$CO/C$^{18}$O abundance ratio. To estimate the $^{13}$CO/C$^{18}$O abundances we first need to estimate $^{12}$C/$^{13}$C abundances for all four clouds. The galactocentric distances for each source lie between 6.5 and 8.5 kpc, which were compared to the $^{12}$C/$^{13}$C gradient measured over the Galactic disk by \citet{langer1990}. This gives a $^{12}$C/$^{13}$C ratio range of between $\sim$ 45-55, and therefore a value of 50 was adopted for the  $^{12}$C/$^{13}$C ratio. Assuming the abundance of $^{16}$O/$^{18}$O in all of the sources to be similar to solar system abundances (i.e. $\sim$ 500; \citealt{zinner1996}), gives a $^{13}$CO/C$^{18}$O abundance ratio of 10.

The gas excitation temperature was estimated using the optically thin C$^{18}$O  line in the following
equation,

\begin{equation}
T_R^*\simeq [T_{\rm{ex}}- T_{\rm{bg}}] \tau_{18} 
\end{equation}

\noindent where $T_R^*$ is the corrected receiver temperature and $T_{\rm{bg}}$ is the background temperature assumed to be $\simeq$ \mbox{2.7 K}. We have assumed the cores are in Local Thermodynamic Equilibrium (LTE) and can therefore be described by a single temperature (i.e. \mbox{$T_{\rm{ex}} = T_{\rm{Kin}} = T$}). The derived core temperature and optical depth are related to the column densities, $N$ (cm$^{-2}$), through the following equation, 

\begin{equation}\label{eq:column_density}
N({\rm{C}}^{18}{\rm{O}})=2.42 \times 10^{14}\tau_{18} \left[ \frac{\Delta v ~T}{1-{\rm{exp}}(-5.27/T)} \right]
\end{equation}

\noindent where $\Delta v$ is the FWHM of the C$^{18}$O line, $\tau_{18}$ and \emph{T} are as previously defined. To convert the C$^{18}$O column densities to H$_2$ column densities a fractional abundance of (C$^{18}$O/H$_2$) = $1.7\times 10^{-7}$ (\citealt{goldsmith1996}) was assumed. The H$_2$ number density was calculated  assuming the cores to be spherical, and that contamination effects from emission along the line of sight can be neglected. The total mass (M$_\odot$) was estimated by multiplying the volume densities of each core by the total volume using,

\begin{equation}
M_{\rm{core}}=6.187\times10^{25}R^3n_{\rm{H_2}}\mu m_{\rm{H}}
\label{eq:mass}
\end{equation}

\noindent where \emph{R} is the radius of the core (pc), \emph{n$_{\rm{H_2}}$} is the molecular hydrogen number density (cm$^{-3}$), $\mu$~is the mean molecular weight (taken to be 2.3, assuming a 25 \% abundance of helium by mass), and \emph{m$_{\rm{H}}$} is the  mass of a hydrogen atom. 
The physical parameters for the molecular cores calculated from the $^{13}$CO and C$^{18}$O data are summarised in Table~\ref{tbl:co_summary5}. We estimate the uncertainties involved in the estimate of the column density to be no more than 50~\%. In estimating the density we have considered the additional uncertainties in the distance to the cores and in the assumptions that the cores are spherical. Taking these additional uncertainties into account we estimate the mass and density calculated to be accurate to within a factor of two.

\begin{table*}
\caption[Summary of parameters derived from $^{13}$CO and C$^{18}$O data]{Summary of parameters derived from $^{13}$CO and C$^{18}$O data.}
\begin{center}

\begin{tabular}{ccccccccc}
\hline
\hline
Cloud	& \vlsr &$\tau_{18}$  & \emph{T}	& Log(N$_{\rm{H_2}}$) & Log(n$_{\rm{H_2}}$)		& Angular  & Physical  	& Mass   \\
id. 		& 	(\kms)	&			& (K)			& (cm$^{-2}$)	 & (cm$^{-3}$) & size (\arcsec) 	& diameter (pc) 				& (M$_{\odot}$) \\
\hline

SFO~58 &3.4  & 0.07 & 37.0	& 22.53 & 4.89 & 47 & 0.16 & 9.5  \\
SFO~68 & $-$16.7&0.06 & 29.4 	& 22.51 & 4.36  & 68 & 0.46 & 66.4 \\
SFO~75 & $-$37.5&0.13 & 23.5    & 22.75 & 4.53  & 41 & 0.56 & 177.3\\
SFO~76 &$-$23.3 &0.05 & 26.7    & 22.26 & 4.06  & 59 & 0.52 & 48.1	\\

\hline
\end{tabular}
\label{tbl:co_summary5}
\end{center}
\end{table*}

The physical parameters of the cores are similar, with the densities ranging between $\sim$ 10$^4$--10$^5$~cm$^{-3}$, the physical diameters varying between $\sim$ 0.2--0.6~pc, kinetic temperatures ranging from $\sim$~24 to 37~K, and masses from $\sim$~10--180~M$_{\odot}$. The temperature of all of the cores are significantly higher than would be expected for starless cores \mbox{(\emph{T} $\sim$ 10 K}; \citealt{evans1999}), which suggests these cores possess an internal heating mechanism, possibly a YSO, or an UC HII region (as indicated by the IRAS colours; see Table~\ref{tbl:IRAS_sources}). It is possible these cores are being heated by the surrounding FUV radiation field, however, if that were the case, we would expect to find a temperature gradient that peaked at the position of the bright rim and decreased with distance into the cloud, but this is not observed. 

\subsection{Physical parameters of the ionised boundary layers}
\label{sect:IBL}
\label{sect:radio_analysis}
Using the radio emission detected toward the rims of SFO~58, SFO~68, SFO~75 and SFO~76 we can quantify the ionising photon flux impinging upon them. Making the assumption that all of the ionising photon flux is absorbed within the IBL we can determine the photon flux, $\Phi$ (cm$^{-2}$ s$^{-1}$), and the electron density, n$_{e}$ (cm$^{-3}$), using the following equations which have been modified from Equations~2 and 6 of \citet{lefloch1997} (see Paper~I for details):

\begin{equation}
\Phi=1.24\times10^{10}S_{\nu}T_e^{0.35}\nu^{0.1}\theta^{-2}
\end{equation} 

\begin{equation}
n_{e}=122.21\times\sqrt{\frac{S_{\nu}T_e^{0.35}\nu^{0.1}}{\eta R\theta^2}}
\end{equation}

\noindent where \emph{$S_{\nu}$} is the integrated flux density in mJy, $\nu$ is the frequency at which the integrated flux density is evaluated in GHz, $\theta$ is the angular diameter over which the flux density is integrated in arc-seconds, $\eta$R is the shell thickness in pc, and \emph{T$_{e}$} is the electron temperature in K. Note, an average HII~region electron temperature of \mbox{$\sim$ 10$^{4}$ K} and \mbox{$\eta$ = 0.2} (\citealt{bertoldi1989}) have been assumed. 

The values calculated for the ionising photon fluxes, and electron densities within the IBLs of the four clouds, are presented in Table~\ref{tbl:physical_parameters}. The main uncertainties in these values are due to flux calibration ($\sim10$~\%), the approximation of the electron temperature, $T_e\simeq10^4$~K (e.g. a difference in the electron temperature of 2000~K corresponds to an uncertainly in the photon flux of $\sim$~25~\%), and the $\eta=0.2$. The uncertainties in flux calibration and electron temperature combine to give a total uncertainty in the calculated photon fluxes and electron densities of no more than 30~\%. However, approximating $\eta=0.2$ leads to the the electron density being underestimated by at most a factor of  $\sqrt{2}$, and since the uncertainty in  $\eta=0.2$ dominates the others it effectively sets a lower limit for the electron density.  

\begin{table*}[!hbt]
\caption[Summary of derived physical parameters of the IBL]{Summary of derived physical parameters of the IBL.}
\begin{center}
\begin{tabular}{cccccc}
\hline
\hline
Cloud & \multicolumn{3}{c}{Photon fluxes (10$^8$ cm$^{-2}$~s$^{-1}$)} & \multicolumn{2}{c}{Electron densities $n_e$(cm$^{-3}$)}\\
id.& Predicted $\Phi_P$ & Peak $\Phi$  & Mean $\Phi$	& Peak  & Mean  \\
\hline
SFO~58 & 20.4 &   31.9 & 19.3 & 338& 262    \\
SFO~68& 126.0 & 68.5 &  32.0  & 341 & 233  \\
SFO~75  &  448.0 & 257 & 40.0 & 839 & 332 \\
SFO~76 & 441  & 263.1&  46.5  & 1242& 526  \\
%\hline
%SFO 64& 19.5 & 48.0 &  23.0   &183 & 133  \\
%SFO 67  & 47.0 & 12.2   &  3.0   &144 & 72  \\
\hline
\end{tabular}
\label{tbl:physical_parameters}
\end{center}
\end{table*}
 
Predicted ionising fluxes were calculated from the Lyman flux of the candidate ionising stars and their projected
distances to the clouds following the method described in \mbox{Paper I}. Comparing the predicted fluxes to the
measured fluxes, and taking into account any attenuation between the OB star(s) and the clouds, we find good
agreement (to within a factor of two). In the majority of cases the measured fluxes at the surface of the BRCs are
lower than the predicted flux, which is to be expected, as the predicted flux is a strict upper limit. The one
notable exception to this is SFO~58 where the measured flux is a factor of one and half times greater than the
predicted upper limit. There are two possible explanations for this discrepancy: the spectral type of the ionising
star is incorrect, or the distance to the HII region is incorrect. We favour the former as a misclassification of
the ionising star's spectral type by even half a spectral class can alter its predicted Lyman flux by up to a factor of two,
which would more than account for the difference in fluxes reported here. 

The high resolution radio observations result in a much tighter correlation between the predicted and measured
fluxes than were found with the low angular resolution data presented in Paper I. The variation between measured and
predicted fluxes for the low resolution data for these four clouds ranged from a factor of a few to more than ten
(in the case of SFO~58). Moreover, the analysis of the distribution of radio emission suggests that two sources
(SFO~68 and SFO~76) lie in the foreground relative to the locations of the ionising star(s), and thus are located
farther from the ionising star(s) than the projected distance, used to derive the predicted flux, would suggest. In
these two cases the correlation could be considerably better than the factor of two quoted above.

The main reason for the improved correlation between the predicted and measured fluxes is because the higher
resolution observations have been able to resolve the radio emission, resulting in the detected emission being much
more tightly peaked. This has allowed more accurate measurements of the flux density to be obtained, and
consequently more realistic values for the ionising fluxes and electron densities to be calculated. Values
calculated from the low resolution observations presented in Paper I suffer due to the large size of the
synthesised beams (typically $\sim$ 90\arcsec~ and $\sim$ 60\arcsec~ for the 20~cm + 6~cm and 13~cm + 3~cm
observations respectively) which dilutes the emission if the IBL is not resolved, resulting in significantly lower
flux  densities being measured. This explains why, in every case, the high resolution radio observations have
resulted in an increase in the calculated ionising fluxes. Another reason is the vast improvement in the \emph{u-v}
coverage obtained by using multiple array configurations, which allows the brightness distribution of the emission
to be more accurately deconvolved from the visibility data. Therefore, although low resolution radio observations
may be useful in identifying clouds that are likely to possess an IBL and are thus subject to photoionisation from
the nearby OB star(s), their main use is to limited to the determination of  global estimates for the physical
parameters of the IBLs.

The mean electron densities calculated for each cloud range between 233--526 cm$^{-3}$, considerably greater than the critical value of \emph{n}$_{\rm{e}}\sim$ 25 cm$^{-3}$ above which an IBL is able to develop around the cloud (\citealt{lefloch1994}). The excellent correlation of the radio emission with the bright-rim of the clouds strongly supports the presence of an IBL at the surface of each of these clouds, confirming their identification as potential triggered star forming regions. It is therefore clear that these clouds are being photoionised by the nearby OB star(s), however, it is not yet clear to what extent the ionisation has influenced the evolution of these clouds, and what part, if any, it has played a part in triggering star formation within these clouds. 

\subsection{Evaluation of the pressure balance}
\label{sect:pressure_balance}
\label{sect:implications_pressure_balance}
In this section the results of the molecular line and radio observations will be used to evaluate the pressure balance between the hot ionised gas of the IBLs, and the cooler neutral gas within the BRCs. Following the method described in Paper~I we calculated the internal ($P_{\rm{int}}$) and external ($P_{\rm{ext}}$) pressures (N cm$^{-2}$) using,

\begin{equation}
P_{\rm{int}}\simeq\sigma^{2}\rho_{\rm{int}}
\label{eq:internal_pressure}
\end{equation}
 
\begin{equation}
P_{\rm{ext}}=2\rho_{\rm{ext}}c^2
\end{equation}

\noindent where $\sigma^2$ is the square of the velocity dispersion (i.e. $\sigma^2=\langle\Delta
v\rangle^2/(8\rm{ln}2$), where $\Delta v$ is the core-averaged C$^{18}$O line width (\kms)), $\rho_{\rm{int}}$ is
the core-averaged density calculated in Section~\ref{sect:co_analysis}, $\rho_{\rm{ext}}$ and $c$ are the ionised
gas density and sound speed (assumed to be $\sim$ 11.4 \kms) respectively. The external pressure term includes 
contributions from both thermal and ram pressure. The electron densities calculated in Section~\ref{sect:IBL}
were used to estimate the ionised gas pressures for each cloud's IBL. The calculated internal and external
pressures are presented in Table~\ref{tbl:pressure_balance}.

\begin{table}
\caption[Summary of the pressure balance analysis]{Evaluation of the pressure balance.}
\begin{center}
\label{tbl:pressure_balance}
\begin{tabular}{ccc}
\hline
\hline
Cloud	& \multicolumn{2}{c}{Pressure ($P/k_B$) ($10^6$ cm$^{-3}$ K)}\\
id.& Internal ($P_{\rm{int}}$)& External ($P_{\rm{ext}}$)\\

\hline

SFO~58 & 5.3 & 7.8\\
SFO~68 & 4.4 & 7.0 \\
SFO~75 & 9.1  & 26.5\\ 
SFO~76 &  1.2 & 15.7\\
\hline
\end{tabular}
\end{center}
\end{table}

The largest two uncertainties in the calculation of the molecular pressure are: the uncertainty in the observed
line temperature, and the possibility that the C$^{18}$O derived density may be affected by depletion, either onto
dust grain ice mantles, or through selective photo-dissociation. The uncertainties in the densities are thought to
be no more than a factor of two, the effects of depletion and photo-dissociation are harder to quantify. However,
all of the core temperatures are considerably larger than 10 K, where depletion is expected to be greatest, and
given that they are embedded within the clouds, away from the ionisation front, where they are shielded from much of the
ionising radiation, these effects are not thought to be significant. We estimate the molecular pressures presented
to be accurate to within a factor of two. The IBL pressures are lower limits (due to the electron densities being
lower limits) and taking account of the uncertainties are considered to be accurate to $\sim$ 30~\%.

As discussed in Section~3.1 theoretical models (\citealt{bertoldi1989,lefloch1994}) have revealed the pressure balance to be a sensitive diagnostic that can be used to determine the evolutionary state of BRCs. Comparing the internal and external pressures calculated from the molecular line and radio observations (presented in Table~\ref{tbl:pressure_balance}),  reveals that all of the clouds are under pressured with respect to their IBLs, and are thus in the process of having shocks driven into them. However, taking account of the possible factor of two uncertainty in our calculations of these parameters, it is possible that two of these clouds are in approximate pressure balance (i.e. SFO~58 and SFO~68); these clouds are likely to be in a post-pressure balance state, and it is therefore possible that any current, or imminent, star formation within these clouds could have been triggered.

The remaining two clouds, SFO~75 and SFO~76, are under-pressured by factors of three and twelve respectively, with respect to their IBLs, strongly suggesting that these clouds have only recently been exposed to the HII region and that shocks are currently being driven into the surface layers of these clouds, closely followed by a D-critical ionisation front. These clouds are likely to be in a pre-pressure balance state where the shocks have not propagated very far into the surface layers; it is therefore unlikely that the molecular cores within these two clouds have been formed by RDI, but are more likely to  pre-date the arrival of the ionisation front and have only recently been exposed to the HII region.  Any current star formation taking place within these clouds is unlikely to have been triggered.

\subsection{Compact radio source associated with SFO~58}
\label{sect:uchii_region}
The 6 cm radio continuum image (see \emph{upper left panel} Figure~3) of SFO~58 clearly shows the presence of a radio source positionally coincident with the $^{13}$CO core embedded within this BRC, both of which lie at the focus of the BRC where the photoionisation induced shock is expected to concentrate the majority of the mass within the cloud. The radio source and embedded $^{13}$CO core are offset from the IRAS point source, but their positional correlation hints at a possible association between the two. At a distance of 700~pc the angular size of the compact radio source ($\sim$~18\arcsec) corresponds to a physical diameter of $<$~0.06~pc, which suggested it might be a compact HII region similar to those found within other BRCs reported in \mbox{Paper I} (i.e. SFO~59, SFO~62, SFO~74, SFO~79 and SFO~85; see \citealt{urquhart2004} for a detailed investigation of SFO~79). 

The radio flux of the compact radio source is consistent with the presence of a single embedded B2--B3 ZAMS star. Furthermore, the correlation of the position of the radio source with that of the molecular core detected in the CO observations, both of which are located at the focus of the bright rim (see Figure~\ref{fig:co_sfo58}), offer some circumstantial support for this hypothesis. However, it is possible that the presence of the radio source is an unfortunate alignment of the cloud with an extragalactic background source. To try to determine the nature of this radio emission we attempted to calculate the spectral index ($\alpha$) of the emission using the integrated flux at both frequencies (i.e. $S_\nu \propto \nu^{\alpha}$), however, this proved inconclusive due to the poor sensitivity of the 3.6 cm map. We therefore tentatively identify this compact radio source as a possible compact HII region embedded within SFO~58. 

\begin{table}
\caption[Physical parameters of CRS 4]{Derived parameters of the compact radio source associated with SFO~58.}
\begin{center}

\begin{tabular}{lc}
\hline
\hline
Position\dotfill & $\alpha$(J2000) = 08:45:26 \\
\dotfill&$\delta$(J2000) = $-$41:15:06\\
Source size\dotfill & $<$ 18\arcsec \\
Physical diameter & $<0.06$ pc \\
3.6 cm flux density\dotfill & 1.69 mJy\\
6 cm flux density\dotfill & 2.28 mJy\\
%Spectral index\dotfill & $-0.55\pm0.44$ \\
log(\emph{N}$_i$)\dotfill & 44.0 photon s$^{-1}$\\
Spectral type\dotfill & ZAMS B2--B3  \\
\hline
\end{tabular}
\label{tbl:embedded_source}
\end{center}
\end{table}

\section{Discussion}

Whilst a full hydrodynamic analysis is beyond the scope of this paper, an insight into the potential effect that exposure to the FUV radiation field has had upon the stability of the cores can be gained from a simple static analysis. Additionally, we will estimate the lifetime of these clouds in the light of the continued mass loss through photo-evaporation by the nearby OB star(s), and evaluate its effect on future star formation within these clouds.

\subsection{Gravitational stability of the cores pre- and post-exposure of the clouds to UV radiation}
\label{sect:stability}

To investigate the effect that the arrival of the ionisation front has on the stability of the cores we need to compare the stability of the cores while they were still embedded within their parental molecular cloud to that of the cores once exposed to the FUV radiation field. In the following analysis we implicitly assume that the cores pre-date the arrival of the ionisation front (as shown in the previous section, this is certainly likely for the cores embedded with SFO~75 and SFO~76), and that there was negligible external pressure from the surrounding molecular material.

The pre-exposure stability of the cores can be estimated using the standard virial equation to derive the virial mass, $M_{\rm{vir}}$ (\msun). Comparing these masses with the core masses calculated from the CO data will give an indication of their pre-exposure stability. The virial mass can be calculated using the standard equation (e.g. \citealt{evans1999}),

\begin{equation}
M_{\rm{vir}}\simeq210R_{\rm{core}} \langle\Delta v\rangle^2
\end{equation} 

\noindent where \emph{R}$_{\rm{core}}$ is the core radius (pc) measured from the integrated $^{13}$CO maps, $\Delta v$ is the FWHM line width of the C$^{18}$O line (\kms). The results are presented in Table~\ref{tbl:virial_mass} with the core masses derived from the CO observations. Comparing the calculated masses of each core with their virial mass, it is clear that three cores were gravitationally stable against collapse while still embedded within their parental molecular cloud, however, taking the errors into account it is possible that SFO~75 was close to being unstable to gravitational collapse. 

Now the effect of exposure to the FUV radiation field of the HII region will be examined. This will be estimated
using a modified version of the virial mass (i.e.~the Bonnor Ebert approach) that takes account of the external
pressure of the surrounding medium, in this case the pressure of the IBL. Following the notation of 
\citet{thompson2004a}, this pressure-sensitive virial mass will be referred to as the \emph{pressurised virial
mass}, $M_{\rm{pv}}$ (\msun), given as, \\ \begin{equation} M_{\rm{pv}}\simeq5.8\times10^{-2}\frac{\langle \Delta
v\rangle^4}{G^{3/2}P_{\rm{ext}}^{1/2}} \end{equation}  \\ \noindent where $P_{\rm{ext}}$ (N m$^{-2}$) is the
pressure of the ionised gas within the IBL, \emph{G} is the gravitational constant, and $\Delta v$ is in \kms. The
calculated values for the pressurised virial masses are presented in Table~\ref{tbl:virial_mass}.

This equation is sensitive to the accuracy of the measured line width, and taking into account the errors involved with Gaussian fits to the spectral lines, the calculated values for the pressurised virial masses are considered to be accurate to within a factor of two (\citealt{thompson2004a}). In this case it was assumed that the external pressure acts over the entire surface of the core, not just the side illuminated by the OB star, and that the cores were pre-existing cores recently uncovered by the expansion of the ionisation front. This is a rather simplistic approach but does allow the effect that exposure to the FUV radiation has on the stability of the cores to be investigated.

\begin{table}
\caption[Summary of core masses: physical, virial and pressurised virial masses]{Summary of core masses as derived from the CO data as well as the virial masses, $M_{\rm{vir}}$, and pressurised virial masses, $M_{\rm{pv}}$.}
\begin{center}
\small
\begin{tabular}{cccccc}

\hline
\hline
Cloud& $M_{\rm{co}}$  & $M_{\rm{vir}}$ & $M_{\rm{pv}}$ & $\dot{M}$ & Lifetime	\\
id.& (M$_\odot$) &(M$_\odot$) &(M$_\odot$) & (M$_\odot$ Myr$^{-1}$) & (Myr)\\
\hline

SFO~58 & 9.5 & 24.2  & 10.7 & 14.9 & 0.64\\
SFO~68 & 66.4  & 255.5  & 152.2 & 58.6& 1.13\\
SFO~75 & 177.3  & 311.1 & 78.2 & 64.6& 2.75\\
SFO~76 &  48.1 & 122.8 & 18.4 & 12.6& 3.80\\

\hline
\end{tabular}
\label{tbl:virial_mass}
\end{center}
\end{table}

Comparing the values for the virial and pressurised virial masses shows that exposure to the FUV radiation field dramatically reduces the mass above which the clouds become unstable against gravitational collapse. The difference between the virial and pressurised viral masses range from a factor of $\sim$~2 (i.e. SFO~58 and SFO~68) to a factor of $\sim$ 7 for SFO~76. Taking account of the errors involved in this analysis, only differences of a factor of four or larger are significant. We are therefore unable to determine if exposure to the FUV radiation has had an impact on the stability of SFO~58 or SFO~68, however, it is clear that the exposure is likely to have had a significant impact on the stability of SFO~75 and SFO~76. Moreover, comparing the pressurised virial masses of the cores with those estimated from the CO data reveals that SFO~75 and SFO~76 are both more than a factor of two more massive than the critical pressurised viral mass. It is therefore likely that the exposure of these  two cores to their respective  HII regions has rendered them unstable against gravitational collapse. However, detailed hydrodynamical or radiative transfer modelling (e.g. \citealt{thompson-white2004,deVries2005}) of higher signal-to-noise molecular line data are needed to investigate the presence of collapse motions in these clouds.

\subsection{Eventual fate of the BRCs}

Once a cloud has reached the cometary stage the shocks dissipate, however, the cloud's mass continues to be slowly
eroded away as the D-type ionisation front continues to propagate into it (\citealt{lefloch1994}). In this
situation the propagation of the ionisation front leads to a constant mass loss in the form of a photo-evaporated
flow into the HII region (\citealt{megeath1997}). The amount of material within the boundary of a cloud is finite,
and thus the effect of the ionisation is to slowly erode the limited reservoir of material available for star
formation. This mass loss eventually results in the total ionisation and dispersion of the cloud, and perhaps the
disruption of ongoing star formation either by disrupting any molecular cores before the accretion phase has begun,
or by  exposing the protostar to the FUV radiation field before the accretion phase has finished. Once the
protostar has been exposed much of the  surrounding envelope of molecular material becomes ionised, therefore
limiting the possible size of the forming protostar (see \citealt{whitworth2004}). Therefore the mass loss rate is
an important parameter that can help determine the effect photoionisation has on current, and future star formation
within BRCs, and in estimating their lifetime.

To evaluate the mass loss we use Equation 36 from \citet{lefloch1994} which relates the mass loss (M$_\odot$~$\rm{Myr}^{-1}$) to the ionising flux illuminating the cloud ($\Phi$ photons cm$^{-2}$ s$^{-1}$), i.e.

\begin{equation}
\dot{M}=4.4\times10^{-3}\Phi^{1/2}R_{\rm{Cloud}}^{3/2}
\label{eqn:mass_loss}
\end{equation} 

The globally averaged photon flux calculated in Section~\ref{sect:radio_analysis} and the cloud radii presented in Paper I were used to estimate the mass loss rate for each cloud using Equation~\ref{eqn:mass_loss}; these values are presented in Table~\ref{tbl:virial_mass} along with an estimate for the lifetime of each cloud. The BRC mass loss rates range between $\sim$ 12--59 $M_\odot$ Myr$^{-1}$, corresponding to cloud lifetimes from as little as $6\times10^5$ yr to several Myr. 

The accretion phase of protostar formation is known to last for several \mbox{10$^5$ yr} (\citealt{andre2000}). Therefore any ongoing, or imminent, star formation within SFO~58, SFO~68, SFO~75 and SFO~76 will be unaffected by the ionisation and mass loss, especially SFO~68 where the star formation already appears to be well developed and unlikely to be affected by the mass loss experienced by the cloud. However, the future star formation within SFO~58 may be adversely affected as the ionisation front propagates into the cloud. Although there is no evidence of any current star formation taking place within either SFO~75 or SFO~76, we have shown these clouds are likely to be undergoing RDI as well as having sufficiently long lifetimes for RDI to be a viable method of triggered star formation.

\subsection{Star formation and the evolution of the BRCs}

Direct evidence of whether star formation within these clouds has been triggered is not readily available, however,
it is possible to investigate the probability that the star formation has been triggered by considering the
circumstantial evidence. It is interesting to note that there is strong evidence for the presence of ongoing star
formation within the two clouds (SFO~58 and SFO~68) that fall into the post-pressure balance cloud category, such
as the molecular outflows (see Section~\ref{sect:co_analysis}), association with OH, H$_2$O and methanol masers
(\citealt{braz1989,macleod1992,caswell1995}, i.e. SFO~68) and the possible association with an embedded UC HII
region (Section~\ref{sect:uchii_region}, i.e. SFO~58). Moreover, the association of the UC~HII region with SFO~58
and of  H$_2$O and methanol masers with SFO~68 --- which are respectively and almost exclusively associated with
Class 0 protostars (\citealt{furuya2001}) and high-mass star formation (\citealt{minier2003}) ---  lead us
to conclude that the star formation within these two clouds is relatively recent, being no more than a few 10$^5$
years old. Contrary to the evidence of recent high-mass star formation within the post-pressure balanced clouds we
find no evidence for any ongoing star formation within either of the two pre-pressure balance clouds (SFO~75 and
SFO~76), short of the presence of the embedded IRAS point source. 

\begin{figure}
\begin{center}
\includegraphics[width=0.49\textwidth]{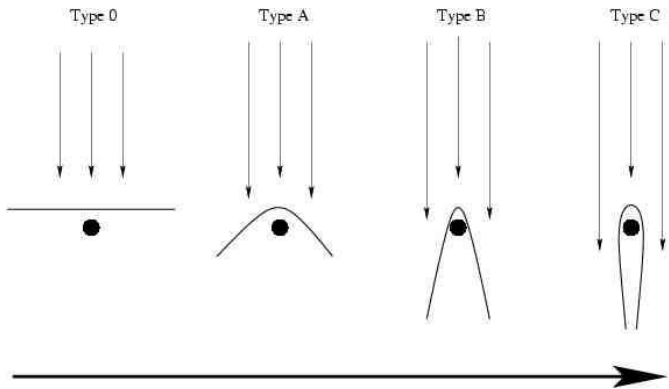}
\caption[BRC rim morphologies]{The above schematic shows the three cloud morphologies, types A, B and C, put forward by \citet{sugitani1991} to classify the BRCs in their catalogue. Another type is added to their original classification scheme; type~0 to represent the morphology of the cloud before being significantly affected by the ionisation front. The thin vertical arrows represent the incoming ionising photon flux and the thick horizontal arrow suggests the possible time-evolution of the clouds under the influence of the FUV radiation field.} 
\label{fig:rim_classification}
\end{center}
\end{figure}

If, as suggested by the RDI models (e.g. \citealt{lefloch1994, vanhala1998}) and observations (e.g. \citealt{sugitani1991}), that rim morphologies represent an evolutionary sequence (see Figure~\ref{fig:rim_classification}), we should expect to find clouds at similar stages of evolution to exhibit similar physical parameters, and furthermore, the star formation within clouds at different evolutionary states to be at different stages of development. It is therefore useful to compare the observational results to the  evolutionary sequence predicted by the models to investigate any differences in the star formation within clouds with different rim morphologies. However, we must point out that the boundary conditions of where the clouds meet the larger-scale molecular material are very important and may affect the following analysis.

To emphasise the morphology of each rim a black curved line has been fitted (by eye) to the radio contours following the minimum gradient of the emission (see \emph{upper right panel} of Figures~\ref{fig:co_sfo58}--\ref{fig:co_sfo76}). The four clouds separate quite nicely into two morphological groups after comparison of the rim morphologies presented in Figure~\ref{fig:rim_classification}. SFO~58 and SFO~68 are typical of a type A rim morphology, whereas SFO~75 and SFO~76 show only slight curvature intermediate between type A and type 0.

Comparing these rim morphologies with those predicted by the RDI models of \citet{lefloch1994} we find that SFO~75
and SFO~76 closely resemble the 0.036 My and SFO~58 and SFO~68 closely resemble the 0.126 My snapshots (i.e. Figure
4a and b of \citet{lefloch1994}). This comparison should be viewed with caution as the Lefloch \& Lazareff models
were calculated for a simple cloud with specific ionising fluxes and so the absolute timescales are more than
likely invalid for our ensemble of BRCs. However the comparison between individual clouds and their relative model
ages does support our conclusion that both SFO~75 and SFO~76 are in the early stages of ionisation, having only
recently been exposed to the ionising radiation of their HII region, and that SFO~58 and SFO~68 have been exposed
for a significantly longer period of time. Moreover, we find that the evolutionary age of the SFO~58 and SFO~68
suggested by the models compares well with the age of the star formation indicator mentioned in an earlier
paragraph (i.e. $\sim$ several 10$^5$ yr). 

Although we have not been able to conclusively prove that the star formation has been triggered within SFO~58 and SFO~68, we have shown that is possible, if not likely. Furthermore, we have shown there are clearly significant morphological and star formation differences between the post- and pre-pressure balance clouds in this survey, which are consistent with the predictions of RDI models. From the observations of the four clouds presented here, it seems that there is reasonably good evidence to support the suggestion that the schematic presented in the Figure~\ref{fig:rim_classification} is an evolutionary picture of the changing morphology of BRCs under the influence of photoionisation.

\section{Summary and conclusions}

In this paper the results of a detailed investigation of four BRCs (SFO~58, SFO~68, SFO~75 and SFO~76) are
presented, including high resolution radio molecular line and continuum observations obtained with the Mopra
millimetre telescope and the ATCA. The main aim is to distinguish between pre- and post-pressure balance clouds,
and to evaluate to what extent the star formation within these BRCs has been influenced by the photoionisation from
the nearby OB star(s). Each of the BRCs was mapped using the $^{12}$CO, $^{13}$CO and C$^{18}$O rotational
transitions using the Mopra telescope. To complement the molecular line observations, high resolution radio
continuum maps of all four BRCs were obtained using the ATCA. 

The CO observations reveal the presence of a dense molecular core within every BRC, located behind the bright rim and, in most cases, coincident with the position of the IRAS point sources. The distribution of the $^{13}$CO emission maps suggest that the cores have a spherical structure and are centrally condensed, consistent with the presence of an embedded gravitationally bound object, such as a YSO. The H$_2$ number densities and cores masses range between 3--$8\times10^4$~cm$^{-3}$ and 10--180~\msun~respectively. The core temperatures ($\sim$ 30~K) are significantly higher than expected for starless cores ($\sim$ 10~K, \citealt{clemens1991}), supporting evidence for the presence of an internal heating source, such as a protostar. 

The high angular resolution radio observations have confirmed the presence of an IBL surrounding the rim of all four
clouds, with the detected emission displaying excellent correlation with the morphology of the cloud rim seen in
the optical images. The increased resolution and improved \emph{u-v} coverage of the radio observations  has
resulted in significantly higher flux densities being measured, which in turn has led to higher values for the
ionising fluxes, which are now more in line with the predicted fluxes than the low angular resolution observations
presented in Paper I. The electron densities are all significantly higher than the critical density of 25~cm$^{-3}$
(\citealt{lefloch1994}) above which an IBL can form and be maintained. All these facts strongly support the
hypothesis that these clouds are being photoionised by the nearby OB star(s). 

CO and radio continuum data are used to evaluate the  pressure balance at the HII/molecular region interface. Comparing these values the clouds are found to fall into two categories: pre- and post-pressure balance states; SFO~75 and SFO~76 are identified as being in a pre-pressure balance state, and SFO~58 and SFO~68 are identified as being in a post-pressure balance state (taking account of the errors, see Section~\ref{sect:pressure_balance}). We draw the following conclusions from our observations:

\begin{enumerate}

\item Analysis has revealed clear morphological and evolutionary differences between the pre- and post-pressure balance clouds:

\begin{itemize}
\item The two clouds identified in this survey as being in a post-pressure balance state are also the same two identified as having a type A rim morphology and show strong evidence of ongoing high- to intermediate-mass star formation (e.g. UC HII region, masers and molecular outflows).

\item The two clouds identified as being in a pre-pressure balance state have all been classified as type 0 clouds, and show no evidence of recent or ongoing star formation.
\end{itemize}

\item The two classifications of rim morphologies, type 0 and type A, correspond to the 0.036 Myr and 0.126 Myr snapshots from the \citet{lefloch1994} RDI model, where the time indicates the exposure time of a cloud to an ionising front. This is consistent with our conclusion that SFO~75 and SFO~76 have only recently been exposed to the ionisation front and that SFO~58 and SFO~68 have been exposed for a significantly longer period of time. Moreover, the morphological age predicted by the RDI models is similar to the estimated age of the star formation within SFO~58 and SFO~68, support the possibility that the star formation has been triggered.  

\item Using a simple pressure-based argument, exposure to the FUV radiation field within the HII regions is shown
to have a profound effect on the stability of these cores. All of the cores were stable whilst embedded in their
natal molecular clouds, however, in all cases exposure to the FUV radiation field of the HII region reduces the
stability of the cores by more than a factor of two (in the case of SFO~76 by almost a factor of seven). The
reduced stability leaves two clouds on the edge of being unstable to gravitational collapse, and two clouds that
have masses at least a factor of two greater than the pressurised virial masses. Analysis of the pre-exposure
stability indicates that it is possible that the core embedded within SFO~75 was unstable to gravitational collapse
prior to being exposed to the ionisation front. From this analysis we conclude that the cores within SFO~75 and
SFO~76 are both unstable to gravitational collapse.  

\item The radio continuum observations toward SFO~58 reveal the presence of an embedded compact radio source within
the optical boundary of the bright rim. The radio source is offset by 1\arcmin~from the position of the IRAS point
source, but correlates extremely well with the position of the peak of the molecular core, both of which are
located at the focus of the bright rim, suggesting that the radio source is associated with the cloud. The size
($<$~0.06 pc) and integrated radio flux are all consistent with the presence of an UC HII region embedded within
the molecular core. The radio flux of this source is consistent with the presence of an UC HII region excited by a
single ZAMS B2--B3 star. Inspection of the core-averaged $^{12}$CO spectrum reveals evidence of a substantial blue
wing, possibly indicating the the presence of a molecular outflow. This is the first tentative evidence for ongoing
star formation within SFO~58.

\item The physical sizes and masses of the molecular cores are typically larger than a single star might be expected to form from. The IRAS luminosities are generally much higher than the bolometric luminosities typical for individual Class 0 and Class I stars (\citealt{andre1993,chandler2000}). Additionally,  evidence that two BRCs are active high-mass star forming regions is presented in this paper, which form exclusively in clusters. It is therefore highly likely that the presence of the IRAS point source indicates the presence of multiple protostellar systems rather than a single protostar. Higher resolution molecular line  observations are required to investigate the possible multiplicity of these sources.

\end{enumerate}

\begin{acknowledgements}
The authors thank the Director and staff of the Paul Wild Observatory, Narrabri, New South Wales, Australia, for their hospitality and assistance during our Compact Array and Mopra observing runs, and the Mopra support scientist Stuart Robertson for his help and advice. We would also like to thank the referee Bertrand Lefloch for his very helpful comments and suggestions. This research would not have been possible without the SIMBAD astronomical database service operated at CDS, Strasbourg, France, and the  NASA Astrophysics Data System Bibliographic Services. We have made use of  Digitised Sky Survey images was produced at the Space Telescope Science Institute under U.S. Government grant NAG W-2166. These images are based on photographic data obtained using the Oschin Schmidt Telescope on Palomar Mountain and the UK Schmidt Telescope. The plates were processed into the present compressed digital form with the permission of these institutions. This research has also made use of the NASA/IPAC Infrared Science Archive,  which is operated by the Jet Propulsion Laboratory, California Institute of Technology, 
under contract with the National Aeronautics and Space Administration.
\end{acknowledgements}

\bibliography{3417}
\bibliographystyle{aa}

\end{document}